\documentclass{ws-ijmpa}
\usepackage[super,compress]{cite}
\usepackage[pdftex,breaklinks]{hyperref}
\usepackage{breakurl}

\def\bs{\boldsymbol}
\def\del{\partial}
\def\bdel{\boldsymbol \partial}

\def\p{{\boldsymbol p}}

\def\k{{\boldsymbol k}}
\def\q{{\boldsymbol q}}

\def\n{{\boldsymbol n}}
\def\x{{\boldsymbol x}}
\def\y{{\boldsymbol y}}

\def\r{{\boldsymbol r}}

\def\u{{\boldsymbol u}}

\def\qqb{{q\bar q}}

\def\tform{t_\text{f}}
\def\tind{t_\text{br}}
\def\kind{k_\text{br}}
\def\Qs{Q_s}
\def\td{t_\text{d}}
\def\tbr{t_{\text{br}}}
\def\kform{k_\text{br}}

\newcommand{\rmd}{{\rm d}}
\newcommand{\beq}{\begin{eqnarray}}
\newcommand{\eeq}{\end{eqnarray}}
\newcommand{\be}{\begin{eqnarray*}}
\newcommand{\ee}{\end{eqnarray*}}
\newcommand{\nn}{\nonumber\\ }

\begin{document}

\markboth{Y.~Mehtar-Tani, J.~G.~Milhano, K.~Tywoniuk}
{Jet physics in heavy-ion collisions}

\catchline{}{}{}{}{}

\title{JET PHYSICS IN HEAVY-ION COLLISIONS
}

\author{YACINE MEHTAR-TANI}
\address{Institut de Physique Th\'eorique,
CEA Saclay, F-91191 Gif-sur-Yvette, France\\
yacine.mehtar-tani@cea.fr}

\author{JOS\'E GUILHERME MILHANO}

\address{CENTRA, Instituto Superior T\'ecnico, Universidade T\'ecnica de Lisboa,\\
Av. Rovisco Pais, P-1049-001 Lisboa, Portugal \and \\
Physics Department, Theory Unit, CERN, CH-1211 Gen\`eve 23, Switzerland\\
guilherme.milhano@ist.utl.pt}

\author{KONRAD TYWONIUK\footnote{Corresponding author.}}
\address{Departament d'Estructura i Constituents de la Materia and Institut de Ci\`encies del Cosmos\\
Universitat de Barcelona,
Mart\'i i Franqu\'es 1, ES-80 028 Barcelona, Spain\\
konrad@ecm.ub.edu
}

\maketitle

\begin{history}
\received{Day Month Year}
\revised{Day Month Year}
\end{history}

\begin{abstract}
Jets are expected to play a prominent role in the ongoing efforts to characterize the hot and dense QCD medium created in ultrarelativistic heavy ion collisions. The success of this program depends crucially on the existence of a full theoretical account of the dynamical effects of the medium on the jets that develop within it. 
By focussing on the discussion of the essential ingredients underlying such a theoretical formulation, we aim to set the appropriate context in which current and future developments can be understood. 

\end{abstract}

\ccode{PACS numbers: 12.38.-t, 24.85.+p, 25.75.-q}


\section{Introduction}
\label{sec:Intro}

Experimental measurements at RHIC\cite{Adams:2005dq,Adcox:2004mh,Arsene:2004fa,Back:2004je} and the LHC\footnote{For a recent review see, e.g., Ref.~\citen{Muller:2012zq} and references therein.} have established extensive and unequivocal   evidence supporting the expectation that a deconfined state of matter, commonly referred to as a {\it quark-gluon plasma} (QGP), is created in the aftermath of ultrarelativistic heavy-ion collisions. 
This hot and dense state provides a unique setting in which dynamic and collective properties of QCD can be studied under conditions resembling those prevalent in the primordial Universe.

Rather than relying on a single `smoking gun' signature, the study of the properties of the QGP resorts to a wealth of complementary measurements which include:
\begin{itemize}
\item the behavior of low-$p_T$\footnote{Throughout, $p_T$ will denote transverse momentum in the detector frame.} hadrons, the bulk outcome of heavy-ion collisions, from which collective properties in both the initial conditions (the saturation of the partonic densities in the colliding nuclei) and the final state (the development of hydrodynamic flows) can be inferred;
\item the modification of the formation probability of quarkonia states due to the presence of a QGP;
\item the modifications, the focus of this review, effected by the QGP on the high-$p_T$ debris produced in the collision.
\end{itemize} 

As realized long ago by Bjorken\cite{Bjorken:1982qr,Bjorken:1982tu},  `hot spots' --- or rather hard partonic collisions --- occur unaffected by the plasma due to the short timescales involved. Thus, the high-$p_T$ products of these processes can be reliably calculated by perturbative methods (see Sec.~\ref{sec:VacJets}). 
The effect of the plasma on these `hard probes', commonly referred to as `jet quenching', and their potential use to extract medium properties have been actively studied for the best part of the last three decades (see Refs. \citen{Appel:1985dq,Blaizot:1986ma,Gyulassy:1990ye,Wang:1991xy} for early pioneering works and Refs.~\citen{Kovner:2003zj,CasalderreySolana:2007zz,dEnterria:2009am,Majumder:2010qh} for recent reviews).

The advent of the LHC heavy ion program --- with a large increase of collision energy (2.76 TeV/c per nucleon pair as compared to the 200 GeV/c available at RHIC) and the excellent detector capabilities of ALICE,  ATLAS and CMS --- have given access not only to a much extended kinematic range for those observables previously explored at RHIC, but also to a whole range of novel measurements involving jets reconstructed reliably and systematically from the midst of the large and fluctuating backgrounds created in nuclear collisions. The availability of these measurements undoubtedly calls for significant refinement of the underlying theoretical descriptions.

Within the limited space available for this survey we, rather than providing an exhaustive overview of developments and results obtained over the last decades (for this purpose we refer the interested reader to the references above), discuss, arguably in a new light, the essential ingredients necessary for quantifying the modifications experienced by jets in ultrarelativistic heavy ion collisions. Early efforts in this direction can be found in  Refs.~\citen{Salgado:2003rv,Vitev:2008rz}.

The aim of the `hard probes' program in heavy-ion collisions --- to extract  information about the properties of the soft and  deconfined QGP --- is pursued via the identification of deviations from a  well-calibrated baseline established in the absence of a medium, i.e. in proton-proton collisions. 
This generic strategy can be illustrated with its application to the case of the nuclear modification factor $R_{AA}$ which evaluates the deviation of single-particle inclusive spectra away from the baseline and is given by
\beq
\label{eq:NuclearModificationFactor}
R_{AB}^X (\{ p_T, b,\ldots\}) = \frac{dN_{AB}^X/d\Omega}{\langle T_{AB} \rangle \, dN_{pp}^X/d\Omega} \,,
\eeq
where $X$ represents the species of the probe and $N_{AB}$ and $N_{pp}$ are, respectively, the inclusive yields in nucleus-nucleus ($A+B$) and proton-proton ($p+p$) collisions ($d\Omega$ is simply the relevant phase space measure). 
The ratio is scaled by the (average) nuclear overlap function $\langle T_{AB} \rangle$, which corresponds roughly to the number of elementary nucleon-nucleon collisions expected for any given centrality class of the nucleus-nucleus collision and is usually extracted from a probabilistic Glauber model\cite{Miller:2007ri,Alver:2008aq}.
The ratio is defined such that $R_{AB} = 1$ when no modification is present. 
The nuclear modification factor is, in general, a function of the collision centrality (equivalently the impact factor $b$), transverse momentum, etc. of the measured particle $X$. 

The same procedure can be used to devise more  general nuclear modification factors that probe two- (and higher-) particle correlations or fully reconstructed jets. Clearly the range of relevant observables is not exhausted by such ratios.  In particular, a significant part of current emerging picture of how jets are modified by the medium has been obtained from observables --- the energy imbalance in dijet pairs\cite{Aad:2010bu,Chatrchyan:2011sx,Chatrchyan:2012nia}, between isolated photons and their recoiling jet\cite{Chatrchyan:2012gt,ATLAS:2012cna}, and between a high-$p_T$ trigger hadron and its recoiling jet\cite{:2012cba}; the geometrical\cite{CMS:2012wxa} and momentum\cite{Chatrchyan:2012gw,ATLAS:2012ina,CMS:2012wxa} fragmentation patterns of jets --- for which information is more readily conveyed through the separate consideration of the measurements in heavy ion collisions and in the baseline than by their ratio.

To attribute the observed modifications, in jet and jet-like hadronic observables, to the presence of a QGP it is essential to account for analogous effects arising from other sources: the nuclear modification of the partonic distributions in the colliding nuclei (an initial state effect); and modifications already present in low-energy nucleus-nucleus and in proton-nucleus collisions,  both cases where no QGP is formed. 
For the high-$p_T$ observables on which this review  is focused,  the nuclear parton distribution functions are well constrained over the relevant kinematical range. 
The effects observed at low energy and in proton-nucleus collisions are due to purely hadronic, essentially non-perturbative,  scattering mechanisms. As such, many of their characteristics --- species (meson/baryon), transverse momentum and rapidity dependences --- are not understood from first principles but are well constrained by existing experimental data.
\cite{Wang:2002ri,Accardi:2007in} 
Generically, these `cold nuclear matter' effects are modest and expected to vanish at high-$p_T$. Indeed, in pPb collisions at LHC energies ($\sqrt{s} = $ 5.02 TeV per nucleon pair) the nuclear modification factor is close to unity for $p_T \geq 5$ GeV/c.\cite{ALICE:2012mj}

Throughout this review we address the interaction of the hard probes with the QGP at the partonic level. Such a perturbative treatment is motivated by the  largeness of the typical probe momentum scales as compared to the ones characterising the medium. 
This separation of momentum scales implies a separation between the timescales related to the probe and the long-distance features of the medium. In other words, the probe (or specific  properties of the probe) is only sensitive to those medium  characteristics involving similar timescales.

One should note, however, that conceptually orthogonal approaches where the probe-medium coupling is taken as large, and necessarily non-perturbative, have been actively pursued in the context of the AdS/CFT correspondence (for a recent and comprehensive review, see Ref.~\citen{CasalderreySolana:2011us}).

In summary, this review addresses what we believe are the essential theoretical ingredients for the description of the development of jets in the presence of hot, dense and colored\footnote{We emphasize that since a hot QGP is a poor color conductor\cite{Selikhov:1993ns}, the partons that propagate through it will exchange color charge at a rapid rate.} medium. 
We restrict ourselves to the discussion of jets initiated by massless partons, that interact perturbatively with the medium, and whose medium modification is dominated by radiative processes.

The many approaches devised to describe medium induced radiation (BDMPS-Z,\cite{Baier:1994bd,Baier:1996sk,Baier:1996kr,Baier:1998yf,Zakharov:1996fv,Zakharov:1997uu,Wiedemann:2000za} GLV,\cite{Gyulassy:2000er} HT,\cite{Guo:2000nz,Wang:2001ifa} AMY\cite{Arnold:2000dr,Arnold:2001ms,Arnold:2002ja} and their extensions) use very similar assumptions about the main features of the process under consideration but differ in the detailed implementation of medium effects and in approximations allowing for analytic treatment. Although these differences play a significant role when comparing to experimental data\cite{Armesto:2011ht},  we refrain from presenting them as competing implementations, but rather attempt to find a common ground from which progress can be made and areas where increased rigor is necessary.

Vacuum jet physics is discussed in Sec.~\ref{sec:VacJets} with particular emphasis on the aspects which will be object of the medium induced modifications addressed in this review. In Sec.~\ref{sec:MediumProp} we describe the medium we take as input for the jet modifications detailed in later sections: medium induced radiation in Sec.~\ref{sec:MediumInducedRad} and  the medium effect on the color properties of a jet in Sec.~\ref{sec:CoherenceMedium}. We conclude in Sec.~\ref{sec:JetQuenchingPhenomenology} with a brief assessment of the phenomenological status of jet quenching.

\section{Jets in vacuum: baseline and template}
\label{sec:VacJets}

\begin{quote}{\it
Starting with the factorization property of QCD cross sections, we discuss in brief the essential features of jet physics in vacuum: 
the collinear nature of gluon radiation and color coherence. }\end{quote}

The essential baseline, with respect to which the modifications of jet observables due to the creation of a QGP can be identified, is provided by the description of the same observables in proton-proton collisions.
In this case, where jets develop in vacuum, the current level of understanding is of enviable maturity and precision.  Jet properties are reliably computed both analytically and in Monte Carlo event generators. 

To a  great extent, the factorizability of short- and long-distance processes in QCD underlies these successes. Such factorizability is evident when the production of a final state $X$ (parton, hadron, jet) is written in the form
\begin{multline}
\label{eq:fsprod}
	\sigma^{h_1\, h_2 \rightarrow X} (p_1,p_2) = f^{h_1}_{i} (x_1,Q^2)\otimes f^{h_2}_{j} (x_2,Q^2) \otimes \sigma^{i\,j\rightarrow k} (x_1p_1, x_2 p_2, Q^2) \\
	\otimes D_{k\rightarrow X}(z,Q^2)\, .
\end{multline}
Here, the PDF $f^{h_{1}}_{i} (x_{1},Q^2)$ ($f^{h_{2}}_{j} (x_{2},Q^2)$) accounts for the probability to find a parton of species $i$ ($j$) within the incoming proton $h_{1}$ ($h_{2}$) carrying a fraction $x_1$ ($x_2$) of the total longitudinal momentum and of virtuality $Q^2$ (the hard scale set by the partonic process). PDFs are universal (process independent) non-pertubative objects with scale $Q^2$ evolution driven pertubatively by the Dokshitzer-Gribov-Lipatov-Altarelli-Parisi (DGLAP) equations.\cite{Gribov:1972ri,Altarelli:1977zs,Dokshitzer:1977sg} The relevant initial conditions are determined via global fits to data and, for the kinematics relevant for jet production, are reasonably  well constrained.

The hard scattering of partons $i$ and $j$, described by the perturbative cross section $\sigma^{i\,j\rightarrow k}$,  results in a partonic system (a pair of back-to-back partons at leading perturbative order) containing the parton $k$ we focus on and which will relax its initial virtuality $Q^2$ through QCD branching down to a scale $Q_0^2$ below which perturbation theory  ceases to be applicable and hadronization of the fragments takes place. 

The probability for the (time-like) parton $k$ (quark or gluon) to branch into partons $l$ and $m$ carrying respectively fractions $z$ and $1-z$ of its momentum is given by
\begin{equation}
\label{eq:splitprob}
	dw^{k\rightarrow l+m} = \frac{\alpha_s}{4\pi}\frac{d^2k_\perp}{k_\perp^2}\,dz\, P_{lk}(z)\, ,
\end{equation}
where the $P_{lk}(z)$ are the Altarelli-Parisi\cite{Altarelli:1977zs} splitting functions which also describe the decay of a negative virtuality, i.e., space-like evolution of PDFs, and $k_\perp$ is the transverse momentum generated in the branching process. 

The well known fact that QCD branching is dominated by the emission of soft gluons at small angles is explicitly seen in Eq.~(\ref{eq:splitprob}): the logarithmic enhancement of small angle radiation from $\int d^2k_\perp\, k_\perp^{-2} \propto \log{Q^2/Q_0^2}$; and the soft enhancement from the integration over $z$ of the contributions in the splitting functions  where a gluon carrying a small momentum fraction is produced ($\propto\int dz/(1-z)$ with $z\rightarrow 1$  and $\propto\int dz/z$ with $z\rightarrow 0$). 

The space-time structure of multiple branchings in the logarithmic regions is characterized by a strong ordering of the typical time scales of successive branchings, $t_\text{f}\sim z(1-z)E/k_\perp^2$.  Unlike the space-like evolution, where the strong ordering in transverse momenta $k_\perp$ accounts fully for the resummation of large logarithmic enhancements, $\alpha_s  \log{Q^2/Q_0^2} \sim 1$, the time-like evolution is determined by a strict angular ordering  of successive branchings due to color coherence effects:

Take the outcome of a branching, a pair of partons with angular separation $\theta$. The radiation of a further parton of energy $\omega$ at angle $ \tilde{\theta}$ from its emitter, in other words a successive branching, takes a finite time which can be estimated from the uncertainty principle as $\tform \sim (\omega \tilde{\theta}^2)^{-1}$.  Since $\omega \tilde{\theta} = k_\perp = \lambda_\perp^{-1}$, with $\lambda_\perp$ the wavelength of the radiated parton, the formation time can be rewritten in the form $\tform \sim  \lambda_\perp/\tilde{\theta}$. Within this time, the original partons separate a distance $ r_\perp \sim \theta\, \tform \sim \lambda_\perp \theta/\tilde{\theta}$.  Hence, for large angle emissions $\tilde{\theta} \gg \theta$, $r_\perp <  \lambda_\perp$,  the radiated quantum cannot resolve the structure of the initial parton pair and probes only its total color charge. Conversely, small angle radiation occurs from the larger color charge of the resolved partons in the pair and is consequently favored.\cite{Mueller:1981ex,Ermolaev:1981cm}

The resumation of logs leads to a set of coupled evolution equations of the Modified Leading-Logarithmic Approximation (MLLA) form\cite{Bassetto:1982ma,Bassetto:1984ik,Dokshitzer:1987nm}
\begin{equation}
\label{eq:dglap} 
\frac{\partial}{\partial \log Q} D_i (x,Q) = \sum_j \int_x^1 \frac{dz}{z} \frac{\alpha_s(k_\perp^2)}{2\pi}\hat P_{ji} (z) D_j (x/z, z Q)\, ,
\end{equation}
where $k_\perp = z(1-z)Q$, for the fragmentation functions $D_i$, where now $\hat P_{ji} (z) $ stands for the regularized splitting functions in the limit $z\to 1$ after the inclusion of virtual corrections.   The leading effects of color coherence are accounted for by the shifted scale $zQ$ in 
$D_j$ on the right-hand side of Eq.~(\ref{eq:dglap}) which effectively implements the angular ordered pattern argued above by restricting the angular range  available for each splitting.
We recall that in standard  DGLAP evolution,\cite{Gribov:1972ri,Altarelli:1977zs,Dokshitzer:1977sg} where color coherence is not  accounted for,  the scale in $D_j$ is simply $Q$.
While unimportant for most values of $x$, this modification is significant at small $x$ and essential for the description of the experimentally observed \textit{humpbacked} plateau in the inclusive energy spectrum of particles in a jet.

Together with the dominance of soft radiation underlying Eq.~(\ref{eq:dglap}), the angular pattern dictated by color coherence defines the space-time structure of jet development: branchings occur successively with decreasing emission angles, the so-called angular ordering of radiation, and are increasingly softer.

So far we have discussed the fragmentation of the initial hard parton $k$ into a partonic system. It is not these  `final state' partons that are experimentally observed, but rather the hadrons they give rise to. The simple hypothesis of Local Parton-Hadron Duality (LPHD)\cite{Dokshitzer:1987nm} which lies in the assumption that the conversion of partons to hadrons affects inclusive observables only via an overall multiplicative factor and the non-pertubative scale $Q_0$ at which perturbative evolution is interrupted, has been very successful; see also \cite{Arleo:2007wn,PerezRamos:2007cr} for calculations of intra-jet $k_\perp$-distributions invoking LPHD. However, for more refined analysis one has to rely on hadronization models to deal with different particles species or for unfolding parton distributions from hadron spectra.

While the, intrinsically non-perturbative, physics of hadronization is not understood from first principles,  it can be effectively encoded by generalizing the fragmentation functions $D_i$. Fragmentation functions $D_{i\rightarrow X}(x,Q)$ describe the probability distribution for a generic  final  state $X$ (hadron, jet, ...) to result from the branchings and subsequent hadronization of fragments of a parton $i$ of which it carries a fraction $x$ of longitudinal momentum. Their evolution is driven by Eq.~(\ref{eq:dglap}) and their non-perturbative input is constrained by global fits to data in a manner analogous to the PDFs. 

Alternatively, as is the case Monte-Carlo event generators, the partonic branching can be performed down to a scale $Q_0 \sim 1$ GeV at which an effective hadronization dynamical procedure is invoked. In these implementations, the partonic fragments are grouped into color neutral structures  (Lund strings \cite{Sjostrand:2006za}, clusters \cite{Webber:1983if}) which dynamically decay into the final state hadrons.

When addressing observables involving reconstructed jets, a strict definition --- an algorithm specifying how to group the fragments completed by a set of parameters (e.g., the jet radius) --- of what the jet is must be given.  Since, in general, different jet definitions result in different jets, comparison between theory calculations and data are only meaningful when the same definition is used. The rationale behind the various existing jet definitions, their applicability and robustness are discussed at length in Ref.~\citen{Salam:2009jx}.

\section{Probing the medium}
\label{sec:MediumProp}

\begin{quote}{\it
The jets probe the underlying medium, which we proceed to discuss. We compare different models for the medium and consider thermal effects that arise in the plasma. Subsequently, the concrete model realization of the medium properties is treated as an input to the calculation of medium effects on jets in the following sections.
}\end{quote}

In the rest frame of a highly energetic particle the incoming medium is strongly Lorentz contracted and nearly translationally invariant. One therefore typically assumes that the probe will not be sensitive to the longitudinal structure of the plasma but only to its static properties. In other words, that the interaction between probe and medium is instantaneous. Leaving a discussion about this point to the end of this section, we will presently implement this approximation which translates into the fact that the momentum exchange is purely transverse. The medium gauge field, $A^-_\text{med}(q) \equiv t^a A^{a,-}_\text{med}(q)$,\footnote{We will work in the light-cone gauge $A^+ = 0$ where $A^-$ is the only relevant component of the medium field.} where $q$ is the momentum transfer from the medium and $t^a$ is a SU(3) matrix in the fundamental representation, can therefore conveniently be written in terms of the effective field
\beq
\label{eq:MediumField}
A^-_\text{med}(q) = 2\pi\, \delta(q^+) \, \int_0^\infty dx^+\, e^{i q^- x^+} A^-_\text{med}({\bs q},x^+) \,,
\eeq
which, in this mixed representation,  depends only on the light-cone time $x^+$\footnote{Light-cone coordinates are defined as $x^\pm \equiv (x^0 \pm x^3)/\sqrt{2}$.} and the transverse momentum $\q \equiv (q_1,q_2)$. To simplify the expression in the following we will identify the symbol $t$ with the light-cone time, $t \equiv x^+$.

In the setup described above, assuming a Gaussian distribution for the medium field, the input from an underlying theory of the plasma enters in the simplest case as a two-point function correlator, which can be written as
\beq
\label{eq:MediumAverage2D}
\langle A^{a,-}_\text{med}(\q,t) A^{\ast \,b,-}_\text{med}(\q', t')\rangle = \delta^{ab} \, n(t) \, \delta(t -t')\, (2\pi)^2 \delta^{(2)}(\q-\q') \, \gamma (\q^2) \,,
\eeq
where $\gamma(\q^2)$ contains the microscopic details of the interaction with the medium constituents and $n(t)$ the density of color charges (which could be a function of the interaction time for expanding media). Note that the correlator is instantaneous reflecting the assumption about a translationally invariant medium. Neglecting higher-order correlators in all observables corresponds to treating the medium as a set of independent scattering centers. In fact, this correlator, at lowest order in the medium coupling $g$, scales with the medium length and is leading compared to higher-order ones in the limit of large media and $g \ll 1$.\cite{DEramo:2012jh}

This can also be understood in terms of screening phenomena in the plasma. Consider the squared Yukawa potential (in momentum space) 
\beq
\label{eq:YukawaScatteringPotantial}
\gamma_\text{GW}(\q^2) = \frac{g^2}{\left(\q^2 + m_D^2\right)^2} \,,
\eeq
which is screened by the characteristic (Debye) mass $m_D$. In coordinate space this corresponds to the fact that the potential extends up to a characteristic distance $r_\text{scr} \sim m_D^{-1}$. This physical setup models the medium as a set of static, randomly distributed scattering centers with a mean free path given by $\lambda_\text{mfp} \sim (n(t) \, \sigma_\text{el})^{-1}$, where the elastic cross section is simply $\sigma_\text{el} \propto \int d^2\q \,\gamma(\q^2)/(2\pi)^2$. This is the so-called Gyulassy-Wang (GW) model of the QGP.\cite{Wang:1991xy,Gyulassy:1993hr,Wang:1994fx} Assuming that $r_\text{scr} \ll \lambda_\text{mfp}$ allows us to treat the scattering centers as independent and justifies the simplifications above. In the opposite case the probe can, in principle, be sensitive to higher-order correlators which capture collective behaviors of the plasma.\footnote{Non-eikonal corrections, allowing for recoil effects, have recently been calculated in Ref.~\citen{Abir:2013ph}.}

Gluon fields at finite temperatures generate screening effects as well. These effects can be studied by high-temperature effective theories, such as the hard thermal loop (HTL) approximation\cite{Braaten:1989mz,Braaten:1990az} (see, e.g., Refs.~\citen{LeBellac:1996xx,Kapusta:2006pm} for reviews). It was found that the longitudinal (electric) gluon fields are screened by a dynamically generated Debye screening mass, which relates to the temperature of the medium as $m_D \sim gT$. The (static) magnetic components are, on the other hand, not screened (see, e.g., Ref.~\citen{Andersen:2004fp} for further details). Since the mean free path scales with an additional factor of the inverse density, and thus scales as $(g^2T)^{-1}$, in the weakly coupled regime, $g\ll 1$, the assumption of independent scattering centers is justified.\cite{Arnold:2002ja}

First and foremost, the interaction with the thermal fields induces thermal masses of the probe due to the modification of their (static) self-energies, see Eq.~(\ref{eq:DerivativeThermalMass}) below. Besides, the probe can alter its kinematics during propagation by exchanging momentum with the thermal medium. Historically, one first calculated the elastic rescattering cross section\cite{Braaten:1991jj,Braaten:1991we} which gives rise to so-called elastic energy loss.\cite{Bjorken:1982tu} Due to the inherent collinear nature of radiative processes it was quickly realized that, although being suppressed by a power of the coupling, they could contribute at the same level as elastic scattering due to phase space enhancement and become dominant for propagating partons of sufficiently high energy. This was first discussed in the context of photon radiation at finite temperatures\cite{Aurenche:2000gf,Aurenche:2002pd,Arnold:2000dr,Arnold:2001ms} and later extended to gluons.\cite{Arnold:2002ja,Besak:2010fb} This gives, in turn, rise to the so-called Landau-Pomeranchuk-Migdal (LPM) effect:\cite{Landau:1953um,Migdal:1956tc} the formation time of induced radiation can exceed the mean free path giving rise to interference effects between subsequent rescatterings, see Sec.~\ref{sec:MediumInducedRad_LPM} for a comprehensive discussion.  Since the radiative processes scale with a larger power of the in-medium path length, see Eq.~(\ref{eq:RadEnergyLoss}) and discussion below, compared to elastic ones, one usually neglects the latter effects for highly energetic probes and large media.\cite{Bjorken:1982tu} While elastic rescattering effects should be incorporated consistently for low-$p_T$ observables,  see also Ref.~\citen{Djordjevic:2006tw} and comment below, we will not currently examine them in more detail.\footnote{Other effects, such as, e.g., transition radiation\cite{Djordjevic:2005nh}, absorptive effects\cite{Bluhm:2011sw,Bluhm:2012kp} and Mach cone creation\cite{CasalderreySolana:2004qm} or Cherenkov radiation\cite{Dremin:1980wx} due to supersonic motion in the plasma, can also play a role but will not be discussed here.}

Then, for soft momentum transfers from the medium, $|\q| \ll T$, the potential (squared) at leading order in the coupling becomes\cite{Aurenche:2002pd}
\beq
\label{eq:HTLScatteringPotential}
\gamma_\text{HTL}(\q^2) = \frac{g^2}{\q^2(\q^2 + m_D^2)} \,,
\eeq
and scales as $\gamma_\text{HTL} \sim \mathcal{N} \q^{-4}$ for $|\q| \gg T$, where the constant is e.g. given in Ref.~\citen{Arnold:2008vd}. Comparing to the static potential, Eq.~(\ref{eq:YukawaScatteringPotantial}), one observes a divergent behavior for small $|\q|$. Higher-order corrections in $g$ to Eq.~(\ref{eq:HTLScatteringPotential}) are also known,\cite{CaronHuot:2008ni} and lead to an even bigger enhancement of the soft sector. Thermal effects are included in several theoretical calculations\cite{Arnold:2000dr,Arnold:2001ms,Arnold:2002ja,Jeon:2003gi,Djordjevic:2007at,Djordjevic:2009cr} of radiative processes in medium, recently also in the presence of a finite chemical potential.\cite{Gervais:2012wd}

\begin{figure}[t!]
\centering
\includegraphics[width=0.65\textwidth]{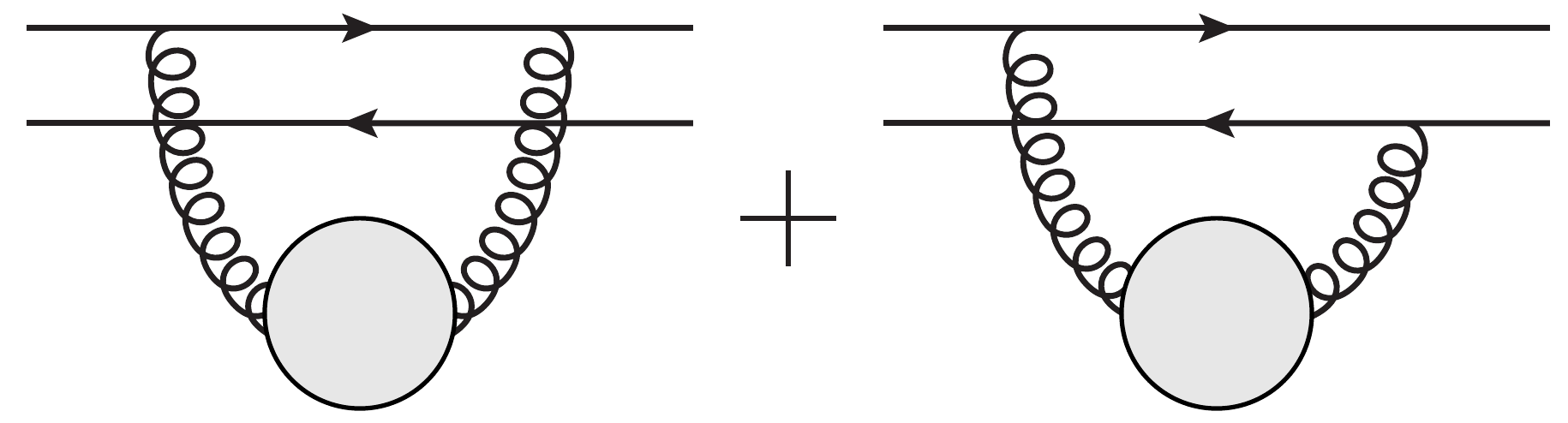}
\caption{Diagrams contributing to the dipole scattering rate in Eq.~(\ref{eq:DipoleCrossSection}) (one must also add the complex conjugate diagrams).}
\label{fig:DipoleCrossSection}
\end{figure}
From our discussion so far, the probe will be sensitive to medium characteristics through interactions which induce dependence on parameters.
The second moment of the correlator in Eq.~(\ref{eq:MediumAverage2D}), historically called $\hat q$, is a measure of the transverse momentum (squared) acquired by the probe per unit length in the elastic scattering and, as we will see below, is a highly important quantity for the study of jets in medium. We will define it stripped of its relevant color factor, as
\beq
\label{eq:QhatDefinitionLocal}
\hat{ q}(t) \equiv \alpha_s n(t)\, \int_{|\q|<q^\ast} d\q^2 \;\q^2 \gamma(\q^2)\,,
\eeq
where $q^\ast$ is a ultraviolet (UV) cut-off and $\gamma(\q^2)$ is given by the medium-model under consideration, cf. Eqs.~(\ref{eq:YukawaScatteringPotantial}) or (\ref{eq:HTLScatteringPotential}).\cite{Arnold:2008vd} Anticipating the detailed discussion in Sec.~\ref{sec:MediumInducedRad}, the relevance of $\hat q$ can be justified by considering how a dipole of transverse size $\r$ interacts via two-gluon exchange with the medium, see Fig.~\ref{fig:DipoleCrossSection}. The resulting scattering rate, stripped of its relevant color factor, is given by
\beq
\label{eq:DipoleCrossSection}
\Gamma_2(\r,t) = 2 g^2 n(t)\, \text{Re} \int \frac{d^2 \q}{(2\pi)^2} \left(1 - e^{i \q\cdot \r} \right) \gamma(\q^2) \,.
\eeq
To leading logarithmic order one finds back $\hat q$ by expanding the term in the brackets to the first non-trivial order and comparing Eq.~(\ref{eq:QhatDefinitionLocal}) with the previous equation it becomes clear that the relevant cut-off $q^\ast \sim |\r|^{-1}$ in Eq.~(\ref{eq:DipoleCrossSection}).\cite{Arnold:2008zu,CaronHuot:2008ni}

For both the static and HTL forms of the potential, Eqs.~(\ref{eq:YukawaScatteringPotantial}) and (\ref{eq:HTLScatteringPotential}), the general form of the scattering rate is $\Gamma_2 \propto \r^2(\log 1/(m_D^2\r^2) +\text{const.})$. In the limit $\r^{-2} \gg m_D^2$ one can neglect the variation of the logarithm which corresponds physically to multiple soft scatterings in the medium. In this limit we can employ the so-called `harmonic oscillator` approximation which reads
\beq
\label{eq:HOapproximation}
\Gamma_2(\r,t) \simeq \frac{1}{2} \hat q(t) \r^2 \,,
\eeq
and has been historically used to ease the analytical treatment of the expressions. The additional logarithmic behavior in the opposite limit arises due to hard momentum transfers with the plasma constituents (quasi-particles).
This situation is in fact highly relevant for highly energetic probes since the UV regulator $q^\ast$ scales with a power of its energy.\footnote{Since $|\r|^{-1} \sim \kform$, see below Eq.~(\ref{eq:HeuristicRadMom}), we expect $q^\ast \propto E^{1/4}$.}\cite{Arnold:2008vd,Arnold:2008zu,CaronHuot:2008ni} This `hard` contribution has a strong impact on phenomenological observables\cite{Armesto:2011ht} and should therefore be treated carefully. 
Keeping this in mind, throughout we will mostly employ Eq.~(\ref{eq:HOapproximation}), commenting on the `hard' dynamics whenever it is relevant.

As will be demonstrated explicitly in Sec.~\ref{sec:MediumInducedRad}, for energetic (eikonal or close-to-eikonal) particles in the medium, the only parameter --- apart from the thermal masses --- that governs the medium-induced dynamics is $\hat{ q}$ defined in Eq.~(\ref{eq:QhatDefinitionLocal}) . It determines the transverse momentum broadening of particles in the medium, see Eq.~(\ref{eq:HObroadening1}), and controls the rate of medium-induced splittings, Eq.~(\ref{eq:BDMPSrate}). This parameter can be calculated from an underlying microscopic theory, e.g., high-temperature QCD in the HTL approximation in isotropic\cite{Arnold:2008vd} or anisotropic\cite{Baier:2008js} background, or on the lattice,\cite{Majumder:2012sh} but it is most usually treated as a phenomenological parameter to be extracted from data.\cite{Bass:2008rv,Armesto:2009zi} 

Above, cf. Eq.~(\ref{eq:QhatDefinitionLocal}), $\hat{q}$ is defined locally and, as such, should be understood as a transport coefficient of the medium. In the literature one has also studied it {\it globally} in the context of the transverse momentum broadening of a probe traversing a medium of length $L$, see also the discussion in Sec.~\ref{sec:MediumInducedRad_Factorization} and Eq.~(\ref{eq:HObroadening1}).\cite{Liu:2006he,DEramo:2010ak,DEramo:2012jh,Benzke:2012sz} Within the approximations discussed above (translationally invariant medium, etc.) these two definitions are equivalent, but the {\it global} approaches opens up the possibility for studying the interplay of medium scales\cite{Laine:2012ht,Benzke:2012sz} and long-distance behavior in more detail. In this case one could also account for the recoil of induced radiation in course of the propagation\cite{Wu:2011kc} in the definition.

Finally, we note that the assumption of a translationally invariant medium should break down for probes with lower energy which subsequently become sensitive to the longitudinal structure of the medium. This gives rise to energy depletion of the probe or so-called {\it drag effects}. In the most general case, the correlator in Eq.~(\ref{eq:MediumAverage2D}) is related to the retarded propagator of the medium fields,\cite{LeBellac:1996xx} and therefore becomes responsive to other moments, i.e. not only involving the transverse momentum components, besides $\hat q$. These novel components are called $\hat e$ and $\hat e_2$ and can be incorporated in analysis of the propagation of a color probe through the medium.\cite{Qin:2012cz} They provide an additional source of energy loss which is  closely related to the elastic energy-loss mechanism mentioned above.

It is worth recalling that the above parametric estimates are valid in a weakly-coupled regime, where $g \ll 1$. Going beyond this approximation is very challenging from a theoretical point of view.\cite{Su:2012iy} In a complementary effort, strong-coupling techniques based on gauge-gravity duality have recently been used to address the problem of energy loss in plasmas; for recent reviews see, e.g., Refs.~\citen{Gubser:2009fc,CasalderreySolana:2011us} and references within. Ultimately, one aims at controlling the remaining dynamics of both probe and heavy-ion system such that one can reliably extract information about medium characteristics on the microscopic level. An interesting perspective in this sense is, e.g., to note the different UV behavior of Eq.~(\ref{eq:DipoleCrossSection}) in weakly and strongly coupled plasmas,\cite{DEramo:2012jh} which arises due to the absence of quasi-particles in the latter regime.

\section{Induced radiation in the medium}
\label{sec:MediumInducedRad}

\begin{quote}{\it
Interactions with the deconfined medium induce the radiation of gluons off the jet. We discuss how this effect probes the medium via the so-called LPM effect and show how this radiation becomes a source of energy loss. Contrary to the collinear nature of radiation in vacuum, this component tends to propagate to large angles and decoheres in color from the remaining jet structure.
}\end{quote}

The suppression of high-$p_T$ particles observed in high-energy heavy-ion collisions was early interpreted in the context of radiative energy loss.\cite{Appel:1985dq,Blaizot:1986ma,Gyulassy:1990ye,Wang:1991xy} These calculations predicted that energetic partons tend to lose their energy via stimulated radiation of soft gluons as a result of interacting with the color charges of the dense medium formed in the collision. The resulting shift in energy, $p_T\to p_T-\Delta p_T$, leads to a depopulation of high-$p_T$ modes. Our goal is to discuss the state of the art of the theory of jet-quenching which is common to most of the present-day calculations, its achievements and limitations, and provide the reader with the basics tools and references.

\subsection{The LPM effect in QCD}
\label{sec:MediumInducedRad_LPM}

Consider an energetic parton of energy $E$, produced in a hard process very early after the collision, typically over a time $1/E$, traversing a medium of length $L$. It will undergo multiple final state interactions with the color charges of the dense medium. Given that the energy of the parton is much higher that the typical momentum acquired due to the multiple scatterings in the medium, i.e., $E \gg \langle p^\text{med}_\perp(L) \rangle$, the parton-medium interactions mainly cause color precession at high rate.\cite{Selikhov:1993ns}  These color excitations stimulate gluon emissions continuously along the in-medium path.

As a na\'ive expectation for the emission rate let us assume that all the $N_{\text{scatt}}=L/\lambda_{\text{mfp}}$ scattering centers along the path act independently. This implies that the spectrum will be an incoherent sum of emissions due to every scattering and reads 
\beq
\label{eq:IndepRad}
\omega \frac{dI^{\text{indep}}}{d\omega} \sim \alpha_s N_{\text{scatt}} \,.
\eeq
This estimate is of course wrong because it ignores coherence effects due to quantum interferences between scattering centers. In fact, medium-induced gluon radiation can be understood in terms of medium resolution power of the higher Fock-states of the high-energy parton. 

As discussed at length in Sec.~\ref{sec:MediumProp}, the medium is characterized by the parameter $\hat q$ which corresponds to the characteristic transverse momentum squared per mean free path, so that $\hat q \sim m_D^2/\lambda_\text{mfp}$ heuristically. A typical fluctuation of the quark state, namely a virtual gluon lives a time $\Delta t$ and can explore the plane transverse to the projectile over a distance $\Delta x_\perp \sim 1/k_\perp$ due to quantum diffusion. During this time, if the average transverse correlation length in the medium $(\hat q \Delta t)^{-1/2}$ --- which is simply the typical inverse transverse momentum that a particle can accumulate during $\Delta t$ --- is larger than the size of the fluctuation, the fluctuation will be transparent to the medium. But as soon as $\Delta x_\perp \sim (\hat q \Delta t)^{-1/2}$, which implies that $k^2_\perp \sim \hat q \Delta t$, the medium resolves the system gluon-quark and the gluon is freed. Making use of the uncertainty principle, $\Delta t \sim k^- \sim \omega/k^2_\perp$, we then obtain an estimate for the characteristic timescale for this process, which we can call the gluon radiation time
\beq
\label{eq:HeuristicRadTime}
\tind = \sqrt{\frac{\omega}{\hat q}} \,,
\eeq
which also gives the characteristic medium-induced momentum 
\beq
\label{eq:HeuristicRadMom}
\kind^2 = \sqrt{\hat q \omega} \,,
\eeq
accumulated during $\tind$. This gives consistently the size of the gluon at formation as $\Delta x_\perp \sim \kform^{-1}$. It follows that gluons traversing the full length of the medium acquire the maximal amount of accumulated transverse momentum, denoted $\Qs = \sqrt{\hat q L}$. The scattering centers involved in the radiation process act coherently as one. Thus, even if the medium is dense the actual in-medium mean-free path of the fluctuation is no longer the elastic one $\lambda_\text{mfp}$ but rather $\tind$ due to color transparency. Therefore the number of effective number of scattering centers is $N_{\text{eff}}=L/\tind$. This qualitative discussion allows us to correct the above naive estimate of the rate of gluon emissions, by replacing $N_\text{scatt} \to N_\text{eff}$ in Eq.~(\ref{eq:IndepRad}), so that
\beq
\label{eq:CohRad}
\omega \frac{dI^{\text{coh}}}{d\omega} \sim \frac{N_{\text{eff}}}{N_{\text{scatt}}}\, \omega \frac{dI^{\text{indep}}}{d\omega} \sim \alpha_s \sqrt{\frac{\omega_c}{\omega}} \,,
\eeq
which involves the characteristic gluon energy $\omega_c \equiv \hat q L^2$ for gluons with $\tind = L$. Thus, coherent medium-induced radiation leads to a suppression of the hard modes scaling as $1/\sqrt{\omega}$ as compared to the incoherent case. This form of the spectrum, Eq.~(\ref{eq:CohRad}), is expected to hold between two critical frequencies $\omega_\text{BH} < \omega < \omega_c$, where we have defined the so-called Bethe-Heitler frequency $\omega_\text{BH} \equiv \hat q \lambda^2_\text{mfp}$. This frequency characterizes the gluons that are radiated over times of the order of the elastic mean free path --- gluons that probe smaller lengthscales are naturally not expected to be induced in the medium. Note that this lower cut-off on energy ensures that the gluons are emitted in the forward direction.\cite{Baier:2006fr} Indeed, demanding that the emission angle is smaller than $\pi/2$, or $\kind < \omega$, leads to $\omega >\omega_\text{BH} > \hat q^{1/3}$, where the last inequality follows from the condition of independent scatterings.

This phenomenon is the QCD analog of the Landau-Pomeranchuk-Migdal\cite{Landau:1953um,Migdal:1956tc,Wiedemann:1999fq} effect in QED that leads to the suppression of the soft photon frequencies, see e.g. Ref.~\citen{Peigne:2008wu}.

\subsection{In-medium branching rate}
\label{sec:MediumInducedRad_Elements}

To see how the physical picture discussed above arises concretely in the calculations it is worth to quickly recall the basis for the techniques that have been developed to calculate radiative processes in medium. In this section we discuss the leading order medium-induced gluon emission.

We shall assume that energies of the quark and the radiated gluon are much larger than their transverse momentum or, equivalently, to the momentum transferred to the medium, so that $p^+\gg p_\perp$ and $k^+\gg k_\perp$.\footnote{The momenta $p$ and $k$ denote the final-state momenta of the projectile and radiated gluon, respectively.} This kinematics corresponds to the so-called eikonal approximation of the projectile-medium interaction in which the field theoretical description becomes equivalent to that of two-dimensional non-relativistic quantum mechanics. Also, in the following we shall restrict our discussion to the purely gluonic case since the generalization to include quark degrees of freedom is straightforward. 

Working in the light-cone gauge, $A^+=0$, we can show that only transverse gluon polarizations propagate, see e.g. Ref.~\citen{Blaizot:2012fh}. In addition, the eikonal approximation implies no spin flip along the trajectory yielding a diagonal propagation of the gluon polarization. Thus,  in the presence of a two-dimensional background field $A^-_\text{med}(\x,t)$, see the left-hand-side of Eq.~(\ref{eq:MediumField}), the gluon propagator ${\cal G}(\x,t ; \y, t')$ between two light-cone times $t$ and $t'$ and transverse positions $\x$ and $\y$, respectively, obeys the following Schr\"odinger-like equation
\beq
\label{eq:2DGreensFunction}
\left[i\frac{\del }{\del t}+\frac{{\bs \del}^2}{2p^+}-igA^-_\text{med}(\x,t)\right] \, {\cal G}(\x,t ; \y, t') = i\delta(t-t')\,\delta(\x-\y) \,,
\eeq
where the transverse position is conjugate to the transverse momentum $p_\perp$ and the light-cone momentum $p^+$ plays the role of a mass and is conserved in the eikonal propagation. Note that in the strictly eikonal limit ($p^+ \gg p_\perp$) the solution for ${\cal G}(\x,t ; \y, t')$ is a Wilson line along the trajectory of the particle, given by $\x(t) = \y + (t-t') \p/p^+$, such that
\beq
\label{eq:2DWilsonLine}
\left. {\cal G}(\x,t ; \y, t')\right|_{p^+\gg p_\perp} = \mathcal{P}_\xi \exp\left[ ig \int_{t'}^t d\xi \, A^-_\text{med} \big(\x(\xi), \xi \big) \right] \,,
\eeq
where $\mathcal{P}_\xi$ signifies path ordering. Non-eikonal corrections in Eq.~(\ref{eq:2DGreensFunction}) reflect additional Brownian motion along the trajectory.

\begin{figure}
\begin{tabular}{c c}
\centering
\includegraphics[width=0.5\textwidth]{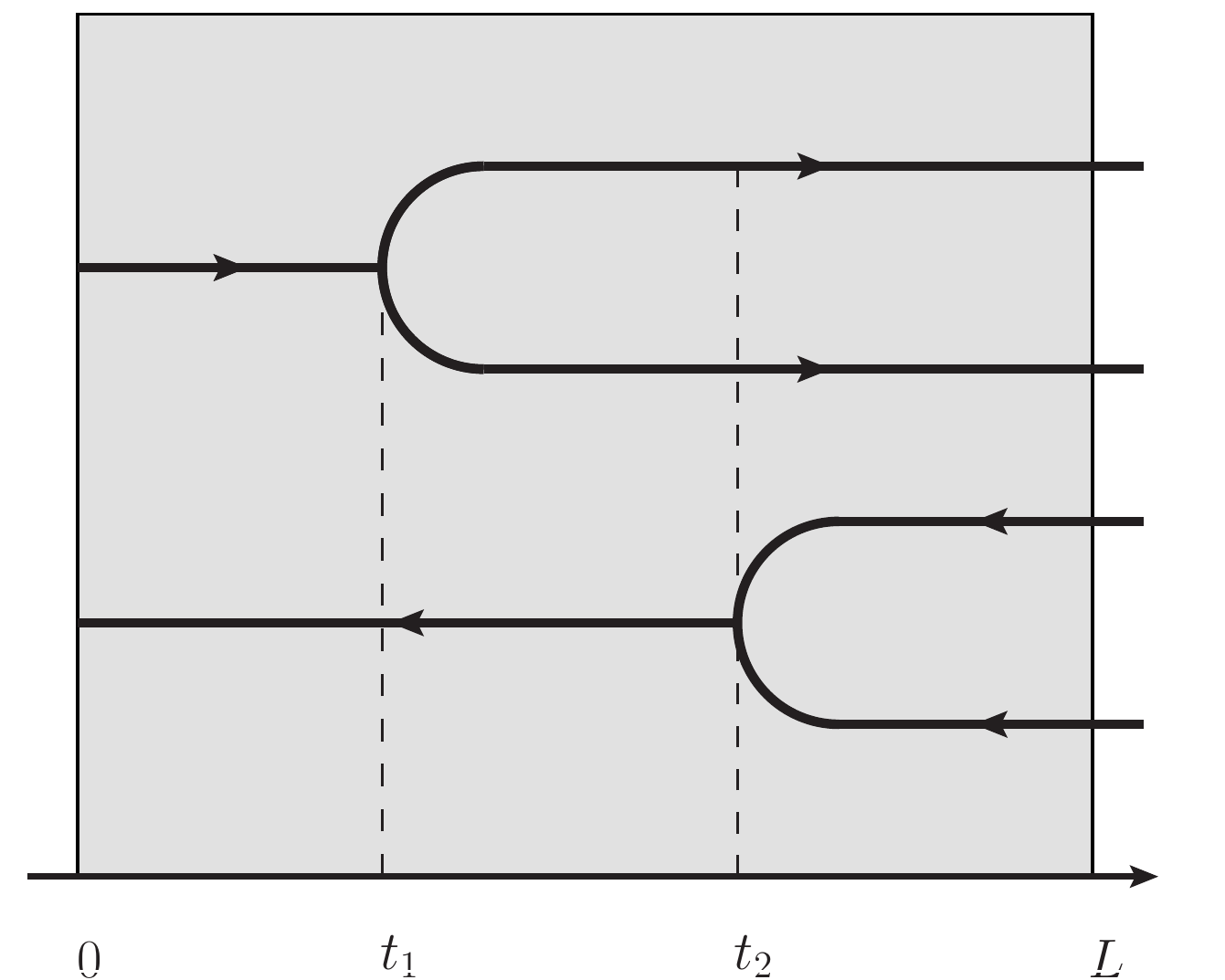} & \includegraphics[width=0.5\textwidth]{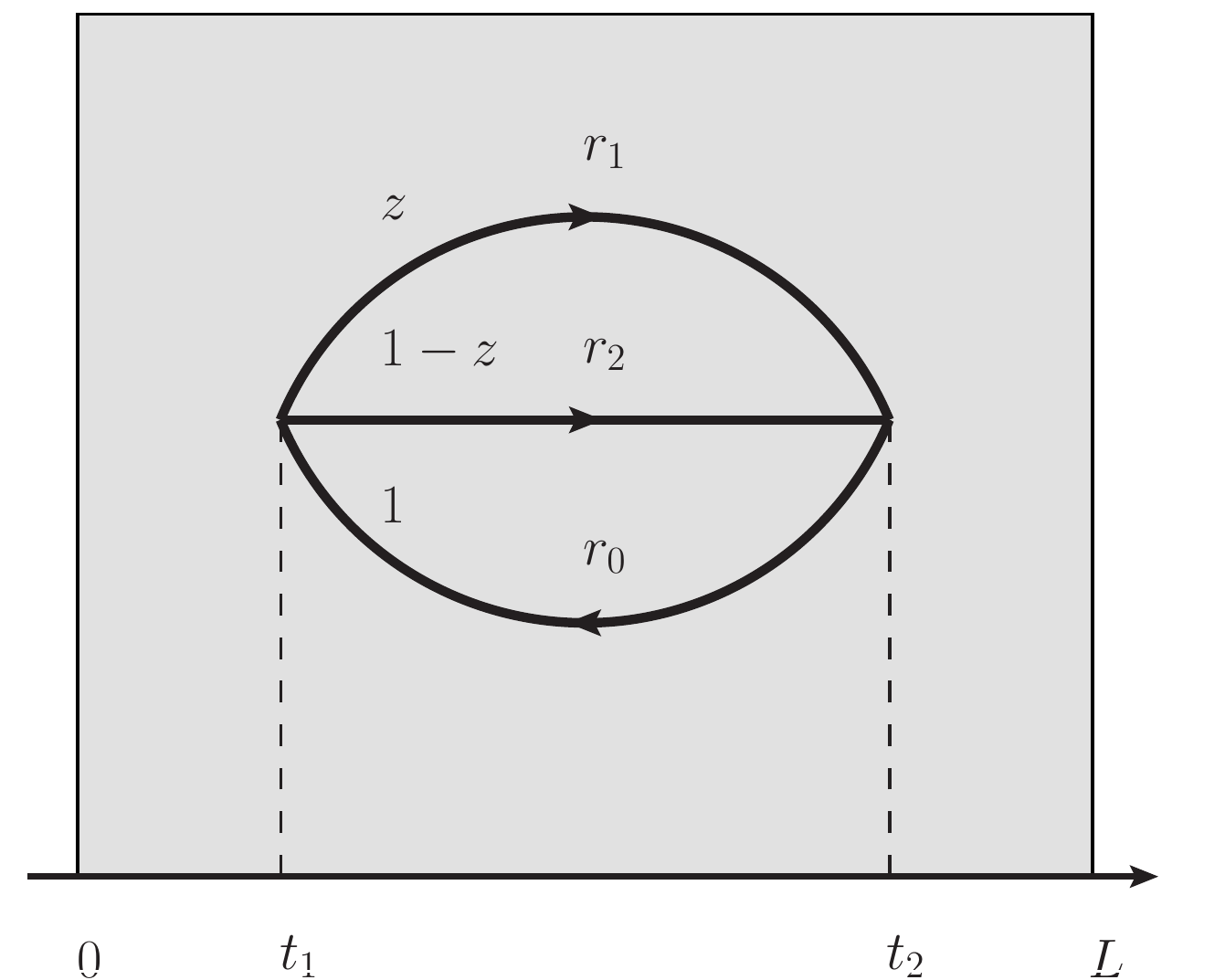}
\end{tabular}
\caption{Left: Square of the gluon emission amplitude in medium. Right: Structure of the squared amplitude after integrating out all transverse momenta.}
\label{fig:EmissionAmplitudeSq}
\end{figure}

Let us now consider the emission of a gluon and for the moment we are only interested in the energy spectrum, postponing the discussion of the transverse momentum dependence to Sec.~\ref{sec:MediumInducedRad_Factorization}.
The square of the amplitude is represented in Fig.~\ref{fig:EmissionAmplitudeSq} (left panel), where the upper lines corresponds to the amplitude and the lower to the complex conjugate. The splitting occurs at the time $t_1$ and $t_2$ in the amplitude and the complex conjugate, respectively. After integrating over the transverse momenta of the outgoing gluons the regions $[0,t_1]$ and $[t_2,L]$ cancel out reflecting the fact that the production process is fully determined by the three-point function between $t_1$ and $t_2$, involving the product of 3 propagators ${\cal G}$, defined in Eq.~(\ref{eq:2DGreensFunction}), in the transverse plane, see Fig.~\ref{fig:EmissionAmplitudeSq} (right panel). The center-of-mass of the system of the three particles, labeled 1 and 2 in the amplitude and 0 in the complex conjugate, is conserved and after integrating over transverse momenta it reduces to zero, i.e., $z\r_1+(1-z)\r_2-\r_0=0$ where $z$ and $1-z$ are the $+$-momentum fraction of the parent parton carried by the offspring partons 1 and 2, respectively. 
As a result, the cross-section will only depend on a single transverse separation $\u$, where $\r_1-\r_0=\u$, $\r_2-\r_0=z\u$ and $\r_1-\r_2=(1-z)\u$.
The medium-induced gluon radiation spectrum reads\cite{Baier:1994bd,Baier:1996sk,Baier:1996kr,Baier:1998yf,Zakharov:1996fv,Zakharov:1997uu}
\begin{align}
\label{eq:BDMPSKernel2-2}
z\frac{dI^{\text{ind}}}{dz} &= \frac{\alpha_s\,zP_{R\to g}(z)}{\left[z(1-z)p^{+}\right]^2}\,\text{Re}\int_{0}^\infty dt_2\int_{0}^{t_2} dt_1\;\nn
&\times \bdel_{\u_1}\cdot\bdel_{\u_2} \Big[ G(\u_2,t_2 ; \u_1, t_1|z)-G_0(\u_2,t_2 ; \u_1, t_1 |z) \Big]_{\u_1=\u_2=0} \,,
\end{align}
which is fully determined by the three-point function $G(\u_2,t_2;\u_1,t_1|z)$ and where $P_{R\to g}(z)$ is the Altarelli-Parisi splitting function for  a parton in a representation R converting into a gluon\cite{Altarelli:1977zs}. It is found after taking the appropriate medium averages, see Eq.~(\ref{eq:MediumAverage2D}), to be given by
\beq
\label{eq:ThreePointFunc_Med}
G(\u_2,t_2 ; \u_1, t_1|z) = \int \mathcal{D}\u \exp \left\{\int_{t_1}^{ t_2} dt  \left[ i\frac{z(1-z)p^+}{2} \dot{\u}^2 - \frac{1}{2}\Gamma_3(\u,t) \right]\right\} \,.
\eeq
This three-point function can also be rewritten in terms of Sch\"odinger equation, which allows for an expansion in terms of the medium density,\cite{Wiedemann:2000za}
\beq
\label{eq:G-eq}
\left[i\frac{\del }{\del t}+\frac{{\bs \del}^2}{2z(1-z)p^+}+\frac{i}{2}\Gamma_3(\x,t)\right] \,G(\x,t ; \y, t') = i\delta(t-t')\,\delta(\x-\y) \,,
\eeq
where the interaction potential of the 3-body system, see Fig.~\ref{fig:EmissionAmplitudeSq} (right panel), reads
\beq
\label{eq:ThreeBodyPotential}
\Gamma_3(\r,t) = \frac{1}{2}C_A \Gamma_2\big(\r,t\big) + \left( C_R-\frac{1}{2}C_A\right) \Gamma_2\big(z\r,t\big) + \frac{1}{2}C_A \Gamma_2 \big((1-z)\r,t \big) \,,
\eeq
where $ \Gamma_2$ is given by Eq.~(\ref{eq:DipoleCrossSection}). In the absence of medium effects, $\Gamma_3 \to 0$, this propagator collapses to the free propagator in vacuum
\beq
\label{eq:ThreePointFunc_Vac}
G_0(\u_2,t_2 ; \u_1, t_1 |z) = \frac{z(1-z)p^+}{2\pi(t_2-t_1)}\exp\left[iz(1-z)p^+\frac{(\u_2-\u_1)^2}{2(t_2-t_1)}\right] \,,
\eeq
which has to be subtracted in Eq.~(\ref{eq:BDMPSKernel2-2}) to obtain the purely medium-induced spectrum.\footnote{A more general regularization prescription was introduced in Ref.~\citen{Wiedemann:2000za}.}

It is worth noting that in the limit of soft gluon emission, i.e. $z \ll 1$, the three-point function collapses to a two-point function which is only sensitive to the broadening of the emitted gluon, since $\r_0 \to \r_2$ in Fig.~\ref{fig:EmissionAmplitudeSq}. In this case, the three-body potential simply picks up the correct color factor for gluon emission, $\Gamma_3(\r) \simeq C_A \Gamma_2(\r)$.

As discussed in Sec.~\ref{sec:MediumProp}, a more detailed analysis of the interaction of the hard parton with a thermal medium can been performed in the framework of the HTL approximation. The main quantitative improvement is the correct treatment of the medium potential entering in $\Gamma_2$ and dressing of the partonic propagators by thermal masses. The latter effect can easily be included in Eq~(\ref{eq:G-eq}) by replacing the derivatives as follows
\beq
\label{eq:DerivativeThermalMass}
\frac{\bdel^2}{z(1-z)p^+} \to \frac{\bdel^2}{z(1-z)p^+} - \frac{m_2^2}{(1-z)p^+} - \frac{m_1^2}{zp^+} + \frac{m_0^2}{p^+} \,,
\eeq
where $m_i$ is the thermal mass of parton $i$ participating in the emission.\cite{Arnold:2000dr,Arnold:2001ms,Arnold:2002ja}

Integrating Eq.~(\ref{eq:BDMPSKernel2-2}) is a difficult task in general and can only be performed numerically,\cite{CaronHuot:2010bp} but one can recover analytical results in some limiting cases. One strategy is to expand the three-point function in powers of the medium density, see Eq.~(\ref{eq:G-eq}). This procedure is commonly called the `opacity expansion`\cite{Gyulassy:2000er,Wiedemann:2000za} and is applicable in the case of a dilute medium or for short emission times. Whenever the number of scattering centers becomes large, one is obliged to resum the contribution from all orders in opacity. 

Another strategy is to pick up the leading-logarithmic behavior, as discussed in the context of Eqs.~(\ref{eq:HOapproximation}) and (\ref{eq:QhatDefinitionLocal}). 
Hence, in the `harmonic oscillator` approximation the integrals over the transverse coordinates become Gaussian and can be performed analytically. Assuming a medium of constant density, $\hat{ q}(t) = \hat{q}_0 \Theta(L-t)$, one finds the well-known BDMPS spectrum 
\beq
\label{eq:BDMPSspectrum}
z\frac{dI^{\text{ind}}}{dz} = \frac{\alpha_s}{\pi} zP_{R\to g}(z) \ln\left| \cos\left[(1+i)\sqrt{\frac{\hat q_{\text{eff}}L^2}{2z(1-z)p^+}}\right]\right|\,,
\eeq
where $P_{R\to g}(z)$ stands for the relevant Altarelli-Parisi splitting function\cite{Altarelli:1977zs} for gluon emission off a quark or gluon, $R = q,g$, respectively, and 
\beq
{\hat q}_{\text{eff}}=\left[(1-z)C_A - z^2 C_R\right]\hat{q}_0 \,.
\eeq
Solutions for a wide class of expanding media are also analytically available,\cite{Baier:1998yf,Arnold:2008iy} see also Refs.~\citen{Salgado:2002cd,Salgado:2003gb} for a general scaling law.

For large media or, equivalently, for gluon emissions meeting the condition $zp^+,(1-z)p^+ \ll \omega_c$, we can simplify the spectrum in Eq.~(\ref{eq:BDMPSspectrum}) further and obtain an emission rate per unit length $dL$, which reads
\beq
\label{eq:BDMPSrate}
\alpha_s\,z{\cal K}(z,p^+)\equiv z\frac{dI^{\text{ind}}}{dz\, dL} = \frac{\alpha_s}{\pi} \sqrt{\frac{\hat q_{\text{eff}}}{2z(1-z)p^+}} zP_{R\to g}(z) \,,
\eeq
which in this approximation is constant. This reflects the fact that gluons can be emitted continuously along the travelled path. Next-to-leading logarithmic corrections have also been found analytically.\cite{Arnold:2008zu} This rate is quite useful for phenomenological applications, see Sec.~\ref{sec:MediumInducedRad_EnergyLoss}.

In the high-energy limit one can neglect the energy of the emitted gluon compared to that of the initial hard parton. This amounts to putting $z\ll 1$, and one typically denotes $\omega \equiv zp^+$. The spectrum in Eq.~(\ref{eq:BDMPSspectrum}) displays two limiting behaviors. When $\omega \ll \omega_c$,
\beq
\label{eq:BDMPSsoft}
\omega\frac{dI^{\text{ind}}}{d\omega}\simeq \frac{2\alpha_s C_{_R}}{\pi}\sqrt{\frac{\omega_c}{2\omega}}\,,
\eeq
which closely resembles our estimate in Eq.~(\ref{eq:CohRad}). It reflects the fact that coherence effects build up when the branching time increases causing the suppression of the spectrum. When $\omega \gg \omega_c $, the radiation time $\tind$ has reached its maximal value $L$ and the spectrum is even more suppressed, namely
\beq
\omega\frac{dI^{\text{ind}}}{d\omega}\simeq \frac{2\alpha_s C_{_R}}{12\pi} \left(\frac{ \omega_c}{\omega} \right)^2\,,
\eeq
Going beyond the leading-log approximation and taking into account the correction from the short distance behavior of $\Gamma_2$ would reduce the power of the suppression to $\sim (\omega_c/\omega)^1$.\cite{Gyulassy:2000er,Wiedemann:2000za,Baier:2001yt} The leading-log approximation is valid as long as $\omega > \omega_\text{BH}$,  where the soft divergence has to be regularized by hand, see the discussion above.

\subsection{Energy loss}
\label{sec:MediumInducedRad_EnergyLoss}
The mean energy of a hard parton with $p^+\gg \omega_c$ is obtained by integrating the BDMPS spectrum, Eq.~(\ref{eq:BDMPSspectrum}), yielding
\beq
\label{eq:RadEnergyLoss}
\Delta E= \int_0^\infty d\omega \,\omega \frac{dI^\text{ind}}{d\omega} \propto \alpha_s C_{_R}\omega_c\,,
\eeq
up to logs of the energy.\cite{Gyulassy:2000er,Baier:2001yt,Zakharov:2000iz} It shows the $L^2$ scaling as opposed to the collisional energy loss that scales as $L$, cf. Sec.~\ref{sec:MediumProp}. Thus, for large enough media radiative energy loss dominates over collisional one. 

In order to compute the quenching of spectra of high-$p_T$ particles one has to go beyond the mean energy loss\cite{Baier:2006fr} and take into account multiple gluon emissions. The rate of emissions, Eq.~(\ref{eq:BDMPSrate}), takes a particularly simple form for soft gluons radiated in the medium. In this limit, $z \ll 1$, we can neglect the requirement of exact energy-momentum conservation and thus it is expected to follow a Poisson distribution. This assumption also relies on the requirement that multiple gluon emissions off the hard parton are independent. While this approximation seems unjustified at present, it is actually valid in a specific kinematic range which will be discussed in the next section. Then, the probability of emitting a total energy $\epsilon$ in course of an arbitrary number of emissions can then be written as\cite{Baier:2001yt,Arleo:2002kh,Salgado:2003gb}
\beq
\label{eq:EnergyLossProbability}
P(\epsilon) = e^{-N_g} \sum_{n=0}^\infty \frac{1}{n!} \, \left[ \prod_{i=1}^n \int d\omega_i \left.\frac{d I^\text{ind}}{d\omega_i}\right|_\text{soft} \right] \, \delta\left(\epsilon - \sum_{i=1}^n \omega_i \right) \,,
\eeq
where the total number of emitted gluons is $N_g = \int d\omega \,dI^\text{ind}/d\omega$. The limits on the $\omega$-integration have already been discussed above. In Mellin space, the summation can be performed analytically, and a solution found for\cite{Baier:2001yt}
\beq
P(\epsilon) = \int_C \frac{d\nu}{2\pi i} \tilde P(\nu) e^{\nu \epsilon} \,,
\eeq
where the contour $C$ runs along the imaginary axis and where the Laplace transform takes a compact form
\beq
\label{eq:EnergyLossProbability_Laplace}
\tilde P(\nu) = \exp\left[-\int d\omega \left.\frac{d I^\text{ind}}{d\omega} \right|_\text{soft }(1-e^{-\nu\omega}) \right]\,.
\eeq
The energy-loss probability defined in Eq.~(\ref{eq:EnergyLossProbability}) has many applications. As an example, let us illustrate the bias toward small energy losses mentioned above defining the so-called quenching factor,\cite{Baier:2001yt}
\beq
\label{eq:QuenchingFactor}
Q^h(p_T) \equiv \frac{d\sigma^{AA\to h+X}/dp_T^2}{\langle T_{AB} \rangle d\sigma^{pp\to h+X}/dp_T^2} = \int d\epsilon \, P(\epsilon) \frac{d\sigma^{pp\to h+X}(p_T+\epsilon)/dp_T^2}{d\sigma^{pp\to h+X}(p_T)/dp_T^2}  \,,
\eeq
which compares the production of hadron $h$ in heavy-ion collisions to proton-proton collisions at the same energy. Since the vacuum spectra typically drop like a power of the transverse momentum $\sim p_T^{-n}$, the support of the integral in Eq.~(\ref{eq:QuenchingFactor}) is biased toward small $\epsilon$. 
Assuming a constant and large slope parameter $n$, the quenching factor is found to scale as $Q^h(p_T) \simeq \exp [-\Delta N(p_T/\pi n)]$, where $\Delta N(\omega)$ is defined in Eq.~(\ref{eq:BDMPSnumber}).\cite{Baier:2001yt} This indicates that only gluons with $\omega < p_T /n$, which is typically smaller than $\omega_c$, contribute to a typical radiative process.

The final distribution of partons with energy $E$ can be evaluated by convoluting the distribution $P$ with an initial distribution, 
\beq
\label{eq:MediumModifiedFragFunction1}
D(E) = \int_0^\infty d\epsilon P(\epsilon)D(E+\epsilon)\,,
\eeq
where the initial condition can, e.g., be taken from perturbative QCD, see Secs.~\ref{sec:CoherenceMedium} and \ref{sec:JetQuenchingPhenomenology} for a further discussion. Furthermore, this distribution obeys the following integro-differential equation\cite{Baier:2000sb,Jeon:2003gi}
\beq
\label{eq:RateEquationSoft}
\frac{\del}{\del L}D(E)=\int d\epsilon \left.\frac{d I^\text{ind}}{d\epsilon \,d L}\right|_\text{soft} D(E+\epsilon) -\int d\epsilon \left.\frac{d I^\text{ind}}{d\epsilon \,dL}\right|_\text{soft} D(E)\,
\eeq
which follows by utilizing the Laplace transform in Eq.~(\ref{eq:EnergyLossProbability_Laplace}).
Without going into details here, it is possible to demonstrate that as long as the gluon emissions are independent this rate equation holds in general for any $z$, so that one can substitute the spectrum in the soft limit by the full rate Eq.~(\ref{eq:BDMPSrate}) in Eq.~(\ref{eq:RateEquationSoft}) and also include mixing between the partonic species.\cite{Jeon:2003gi,Turbide:2005fk,Schenke:2009gb} In addition to parton splittings, this rate equation also includes absorption of thermal particles. In the general case, $P(\epsilon)$ is not anymore described by the Poisson ansatz as in Eq.~(\ref{eq:EnergyLossProbability}). In the following subsection we will indeed demonstrate that this assumption can be proven in a well-controlled way.

\subsection{Factorization of multiple gluon branchings}
\label{sec:MediumInducedRad_Factorization}

From Eq.~(\ref{eq:BDMPSsoft}) one can estimate the average number of 
gluons emitted with energies larger than a given value $\omega$,
 \beq
 \label{eq:BDMPSnumber}
 \Delta N(\omega) = \int_{\omega}^\infty d\omega'\, \frac{d I^\text{ind}}{d \omega'}
 \,\sim\,\bar\alpha \,\frac{L}{\tind}\,,\eeq
 where $\bar\alpha\equiv \alpha_sC_R/\pi$. As long as $\Delta N(\omega)\lesssim 1$, i.e. $\omega \gtrsim \bar\alpha^2\,{\omega_c}$, it may be identified as the probability to emit one gluon with energy $\omega' \ge \omega$. This is the case for the relatively hard emissions that dominate the energy loss, see Eq.~(\ref{eq:RadEnergyLoss}). In this regime the probability of multiple emissions is small. But for sufficiently soft gluons, such that
\beq
\label{eq:SoftGluonMultiple}
\omega < \bar\alpha^2 \omega_c \,,
\eeq
$\Delta N(\omega)$ becomes larger than unity and multiple emissions are clearly important.\cite{Blaizot:2012fh}

In technical terms, this means that all powers of $\bar\alpha L/ \tbr$ have to be resummed. By contrast, the soft and collinear divergences of vacuum radiation give rise to the logarithmic enhancement in terms of the jet energy, $\alpha_s \log^2 E$. We ignore this for the time being by formally setting $L\to \infty$, and postpone a discussion of this issue to Sec.~\ref{sec:CoherenceMedium}.

The treatment of multiple emission is a priori complicated due to interferences between various higher-order processes. For instance, for radiation in vacuum these interferences are essential and give rise to angular ordering, see Sec.~\ref{sec:VacJets}. This mostly affects the emissions of soft gluons which typically have long formation times, $\tform \sim 1/\theta^2\omega$. On the contrary, soft gluons induced by the medium are produced very rapidly, see Eq.~(\ref{eq:HeuristicRadTime}). In fact, for the regime of multiple emissions, defined by Eq.~(\ref{eq:SoftGluonMultiple}), we naturally obtain that $\tbr \ll L$ --- allowing us to treat these emissions as quasi-instantaneous. Below, we will also show that $\tbr$ represents the time when the products of the branching decohere in color,\cite{Blaizot:2012fh} see Eq.~(\ref{eq:FourPointFactorization}).\footnote{For a more general discussion of this physics, we refer to Sec.~\ref{sec:CoherenceMedium}, see in particular Eq.~(\ref{eq:DecoherenceTime}).} In other words, after a time $\sim \tbr$ the partons formed in the branching propagate independently of each other and all interferences are suppressed, see Sec.~\ref{sec:CoherenceMedium}. 

Consider two successive branchings. The smallness of $\tbr$ severely limits the available phase space for interferences, which scale like $\Delta N^{\text{int}}\propto (\alpha_s L) (\alpha_s \tbr)$ due to the requirement of overlap close to the branching. The factorizable piece, on the contrary, scales with the maximal phase space, $\Delta N^{\text{fact}}\propto(\alpha_sL)^2$, due to the quasi-instantaneous nature of the radiation. In line with the discussion above, it follows that the interferences are suppressed as compared to the factorizable contribution when $\omega\ll\omega_c$ by the factor
\be
\frac{\Delta N^{\text{int}}}{\Delta N^{\text{fact}}}\sim \frac{ \tbr}{L}\ll 1\,.
\ee
In order to address the issue of multiple branchings in more detail in the regime discussed above we shall proceed by analyzing the two first orders in a parton cascade which will allow for a generalization to all orders. We shall also extend the discussion of the previous section by discussing quantities that are fully differential in transverse momenta.

Consider a gluon produced at an initial time $t_0$ by a hard process described by the cross section $d\sigma_\text{hard}/d\Omega_{p_0}$, where $d\Omega_k\equiv (2\pi)^{-3} d^2\p_0 d p_0^+ /2p_0^+$ is the standard phase space measure. The cross section for observing the gluon at some later time $t_L$ with transverse momentum $\k$ is obtained by convoluting the initial spectrum at time $t_0$ with the probability for propagation from $t_0$ to $t_L$ which is described by a similar evolution equation as Eq.~(\ref{eq:G-eq}) with $\Gamma_3$ replaced by $C_A \Gamma_2$.
The result reads
\beq
\label{eq:Sigma0}
\frac{d\sigma_0}{d\Omega_k} = \int \frac{d^2\p_0}{(2\pi)^2} {\cal
P}(\k-\p_0; t_L,t_0)~\frac{d\sigma_{\rm hard}}{d \Omega_{p_0}},
\eeq
where ${\cal P}(\k-\p_0;t_L,t_0)$ represents the probability that the gluon acquires a transverse momentum $\q\equiv\k-\p_0$ from the medium in course of the propagation. This is the first building block of our probabilistic picture for jet evolution. In general it is given by 
\beq
\label{eq:HObroadening1}
{\cal P}(\q; t_L,t_0) = \int d^2\r\,\exp\left[-i\q\cdot\r - \frac{1}{2}C_A\int_{t_0}^{t_L}dt'\, \Gamma_2(\r,t')\right]\,.
\eeq
For homogeneous media and in the ``harmonic oscillator" approximation, valid as long as $|\q| \ll \hat q(t_L-t_0)$, it takes a particularly simple form
\beq
\label{eq:HObroadening2}
{\cal P}(\q,\Delta t)\simeq \frac{4\pi}{\hat q \Delta t } \, \exp\left(-\frac{\q^2}{\hat q \Delta t }\right)\,,
\eeq
where $\Delta t =t_L - t_0$. In the opposite regime of `hard` interactions with the medium, see discussion around Eq.~(\ref{eq:HOapproximation}), this distribution behaves as ${\cal P}(\q,\Delta t )\sim \q^{-4}$. This classical propagation in the medium is characterized by the global scale $\Qs^2 = \hat q L$, corresponding to he maximum transverse momentum that a particle can acquire since typically $\Delta t \sim L$. Thus, due to the broadening in medium the propagating particle tend to deflect to large angles. See also Sec.~\ref{sec:MediumProp} for a further discussion of the role of $\hat q$.

\begin{figure}
\centering
\includegraphics[width=5cm]{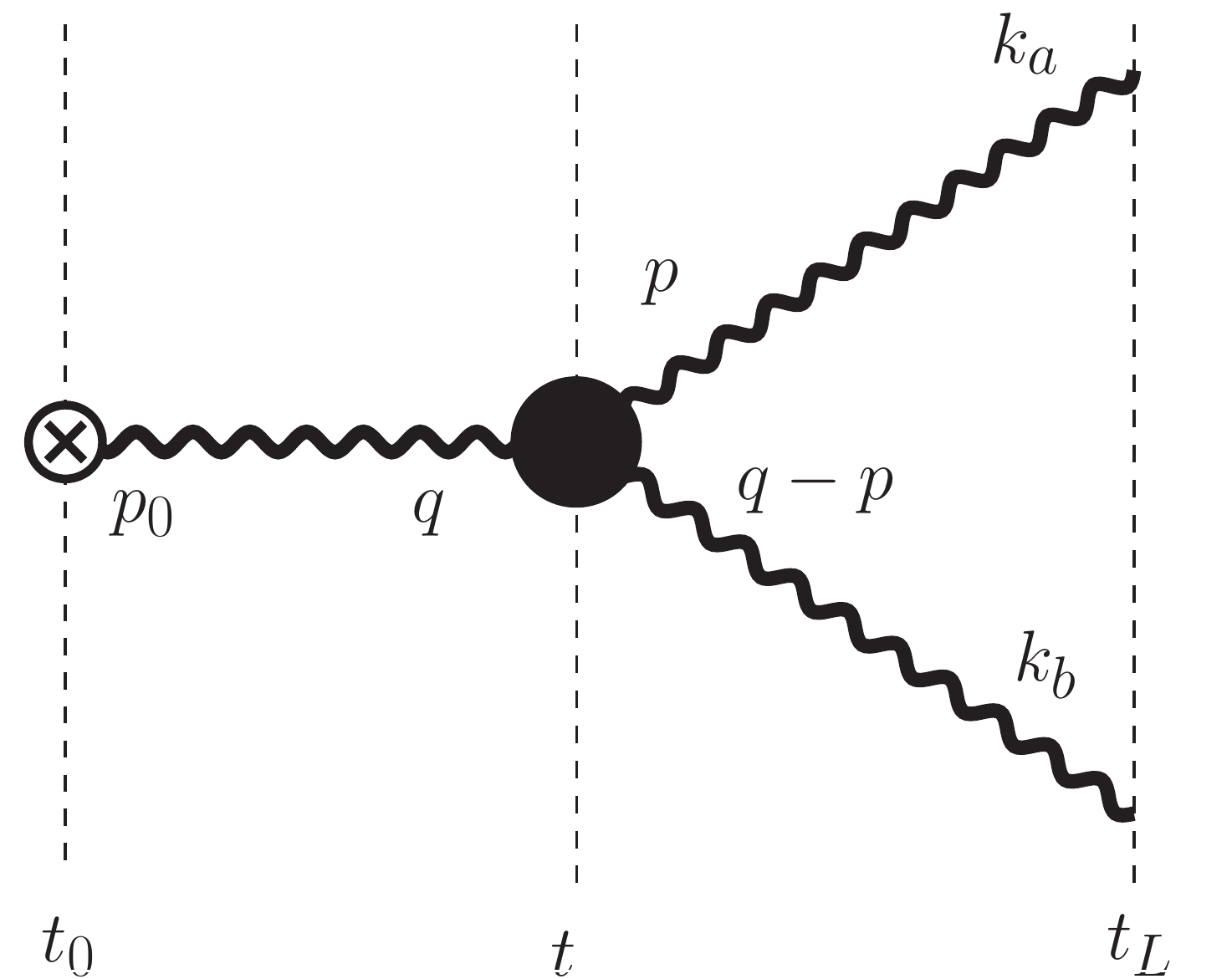}\\
\caption{Graphical illustration of the equation (\ref{eq:Sigma1a}).
The thick wavy lines represent the probability ${\cal P}$ for transverse
momentum broadening,  the black dot is the splitting probability
${\cal K}$, and the circled cross is the cross section of the hard
process producing a gluon of momentum $p_0$. }
\label{fig:GluonBranching}
\end{figure}

We now turn to the cross section for the splitting of an initial gluon into two offspring gluons, labelled $a$ and $b$. This process is depicted in Fig. (\ref{fig:GluonBranching}). We distinguish three time intervals: (i) the propagation before the splitting, described by Eq.~(\ref{eq:Sigma0}); (ii) the splitting process itself, described by a three-point function, as discussed in Sec.~\ref{sec:MediumInducedRad_Elements}; and (iii) the subsequent propagation of the produced pair.

Analogously to Sec.~\ref{sec:MediumInducedRad_Elements}, it can be shown that emission region (ii) has support at most of the order of the branching time $\tbr = \sqrt{z(1-z)p^+_0/\hat q_\text{eff}}$.\cite{Blaizot:2012fh} Since we are working in the limit $\tbr \ll L$, this region is replaced by a quasi-local vertex in Fig.~\ref{fig:GluonBranching} describing the quantum branching process. 

After the products of the branching are formed, their further propagation in region (iii) involves their correlations in color and momentum space and is in general described by an, a priori unknown, four-point function $S^{(4)}(t_L,t)$. The key result of the factorization proof,\cite{Blaizot:2012fh} is to demonstrate that the four-point function factorizes into independent propagations of the two gluons in the limit $\tbr\ll L$ up to corrections of the order of $\tbr/ L$

\beq
\label{eq:FourPointFactorization}
S^{(4)}(t_L,t)\propto {\cal P}(\k_a-\p,t_L-t){\cal P}(\k_b-\q+\p,t_L-t)+{\cal O}\left(\frac{\tbr}{L}\right).\label{S4tildefac}
\eeq
This proves the decoherence of induced gluons shortly after they are produced, i.e. at times larger than $\tbr$. The resulting spectrum at leading order in $L/\tbr$ reads then\cite{Blaizot:2012fh}
\begin{align}
\label{eq:Sigma1a}
\frac{d^2\sigma}{d\Omega_{k_a} \,d\Omega_{k_b}} &= 2g^2 z(1-z)\int_{t_0}^{t_L}\,\rmd t\,\int_{\p_0, \q,\p} \; {\cal P}(\k_a-\p,t_L-t)\, {\cal P}(\k_b-\q+\p,t_L-t)\nn
& \times  \,{\cal K}(\p-z\q,z,p_0^+,t) \,
 {\cal P}(\q-\p_0,t-t_0)\, \frac{\rmd\sigma_{hard}}{\rmd\Omega_{p_0}}\,,
\end{align}
where $z=k_a^+/p_0^+$ and we have adapted the notation $\int_{\p} = \int d^2 \p/(2\pi)^2$. This result can be interpreted as a classical branching process, as illustrated in Fig.~\ref{fig:GluonBranching}, in the following sense. After propagating from $t_0$ to $t$, during which it acquires a transverse momentum $\q-\p_0$, the original gluon splits into offsprings $a $ and $b$ with a probability $\sim\alpha_s {\cal K}(\p-z\q,z,q^+)$ which depends on the longitudinal momentum $q^+$ of the parent parton, the longitudinal momentum fraction $z=p^+/q^+$ carried by gluon $a$,  and the transverse momentum difference  $\p-z\q$.\footnote{The conservation of longitudinal momentum implies of course $p_0^+=q^+=k_a^++k_b^+$ with $k_a^+=p^+=zq^+$.} After the splitting, the two gluons $a$ and $b$ continue to propagate through the medium, from $t$ to $t_L$, 
thus acquiring additional transverse momentum. 

The quasi-instantaneous, $k_\perp$-differential splitting kernel can be computed, similarly as in Sec.~\ref{sec:MediumInducedRad_Elements}, yielding
\begin{align}
\label{eq:SplittingKernel3}
{\cal K}(\p,z,p_0^+,t) & =\frac{P_{gg}(z)}{\left[z(1-z)\,p^+\right]^2}\,\text{Re}\int_{0}^\infty d\Delta t ~ e^{-i\u_2\cdot \p}\;\nn
& \bdel_{\u_1}\cdot\bdel_{\u_2} \left[ \,G(\u_2,t +\Delta t; \u_1, t,z)-G_0(\u_2,t+\Delta t ; \u_1, t,z)\right]_{\u_1=0}\,,
\end{align}
cf. Eq.~(\ref{eq:BDMPSKernel2-2}). For a homogeneous medium the splitting kernel is independent on the emission time $t$. One can go further an evaluate Eq.~(\ref{eq:SplittingKernel3}) in the ``harmonic oscillator" approximation, where it reads\cite{Blaizot:2012fh}
\beq
\label{eq:SplittingKernel3HO}
{\cal K}(\p,z,p_0^+) \simeq \frac{2}{z(1-z)p^+_0} 
P_{gg}(z)\, \sin\left(\frac{\p^2}{2 \kform^2}\right)\,\exp\left(-\frac{\p^2}{2 \kform^2}\right) \,,
\eeq
where $\kform^2=\sqrt{z(1-z)p^+_0 \hat q_\text{eff}}$ is the typical transverse momentum generated during the branching process.
This kernel generalizes the result obtained in Ref.~\citen{MehtarTani:2012cy} in the eikonal limit. 

The spectrum in Eq.~(\ref{eq:Sigma1a}) established the factorization of induced gluon radiation in medium and constitutes the building block for jet evolution in the limit $\tbr\ll L$ at large angles, away from the jet core. Note that integrating Eq.~(\ref{eq:SplittingKernel3HO}) over the transverse momentum we recover the branching rate defined in Eq. (\ref{eq:BDMPSrate}).  Hence, the independent branching approximation assumed in the rate equation introduced in \cite{Baier:2000sb,Jeon:2003gi} is justified so long as the typical in-medium branching time is much smaller than the medium length and can be generalized by including the transverse momentum dependence of the in-medium shower \cite{Blaizot:2012fh}.

\section{Color coherence and color flow in the medium}
\label{sec:CoherenceMedium}

\begin{quote}{\it
Having discussed the main elements of the physics of jet evolution in vacuum, on one hand, and the dynamics of medium-induced radiation, on the other, we can now combine these insights to study how the intrinsic jet-structure is modified by the QGP. For the sake of simplicity, these aspects will mostly be discussed in the context of the smallest building block of the jet, the antenna system, where one most easily can illustrate the effects of color decoherence.
}\end{quote}

So far we have discussed aspects of the branching of a highly energetic particle separately in vacuum and in medium (where in the former case the particle must be strongly off-shell to be able to emit). The complications arising from resumming multiple gluon emissions were treated in two completely different manners in the two cases. In the former case, the probabilistic interpretation of the shower at leading-logarithmic approximation is saved by the condition of angular ordering of subsequent radiation, see Sec.~\ref{sec:VacJets}. In the latter, on the other hand, at least in the regime of soft gluon radiation, i.e. $\omega < \omega_c$, one can rigorously prove the complete factorization of subsequent radiation in the medium, see Sec.~\ref{sec:MediumInducedRad_Factorization}. Thus, while a vacuum jet is a completely color coherent object the shower created via medium-induced branchings is completely decoherent. The natural evolution variable of the two types of shower is also different: in the vacuum the evolution is ordered in virtuality,\footnote{Equivalently, in transverse momentum or angle.} see Sec.~\ref{sec:VacJets}, while in the medium the emission rate of independent gluons scales with the (physical) length of the medium, Eq.~(\ref{eq:BDMPSrate}). 

Dealing with jet systems with an intrinsic scale much larger than any scale related to the medium, e.g. $E\equiv p_T \lesssim 300$ GeV/c at LHC, we should therefore proceed to discuss these seemingly disparate physical processes in a common framework.  A complete theory of jets in heavy ion collisions would be dictated by the interplay of the relevant hard scales in the problem. While the dynamics of jets  in vacuum is fully determined by the transverse mass of the jet $Q=E\Theta$, where $\Theta$ is the jet opening angle, and the non-perturbative scale $Q_0$, in the presence of a medium of length $L$  two additional global scales arise: the typical transverse momentum acquired by a particle traversing the medium $\Qs=\sqrt{\hat q L}$, which inverse relates to the resolution power of the medium to color charges in the transverse plan, and the typical   inverse size of the jet in the medium $r_\perp^{-1}=(\Theta L)^{-1}$.  As we shall discuss in what follows, the competition between these two scales determines the degree of alteration of color coherence of the parton shower.

Naturally, in addition to inducing radiation,  the presence of a deconfined medium --- i.e., a medium interacting with the probes via color exchanges --- is also expected to alter the coherent structure of the jets. Thus, the notion of a resolution scale of the medium needs to be established. This was first studied in the context of radiation off a color correlated pair of partons, the so-called {\it antenna} configuration,\cite{MehtarTani:2010ma,MehtarTani:2011tz,MehtarTani:2011jw,Armesto:2011ir,CasalderreySolana:2011rz,MehtarTani:2011gf,MehtarTani:2012cy} traversing the QGP.\footnote{Note that the color charge of the pair is not necessarily balanced, so that the antenna can possess a total charge.} 

For two color-correlated probes propagating through the medium with momenta $p_1$ and $p_2$, the probability to lose their color-correlation after a certain (light-cone) time $t$ prior to further gluon radiation, is described by \cite{MehtarTani:2011tz,CasalderreySolana:2011rz,MehtarTani:2012cy}
\beq
\Delta_\text{med}(t) 
= 1- \exp\left[-\frac{1}{2} \int_0^t dt' \, \Gamma_2(\delta \n_{12} t',t') \right] \,,
\eeq
where $|\delta \n_{12}| \sim \theta_{12}$ is the angle between momenta of the constituents and $\Gamma_2(\r,t)$ is given by Eq.~\ref{eq:DipoleCrossSection}. We have assumed that the particles are both energetic enough so that the second term in Eq.~(\ref{eq:2DGreensFunction}) can be neglected. In the `harmonic oscillator` approximation this expression takes a particularly simple form, involving only the characteristic hard scales of the problem, namely
\beq
\label{eq:DeltaMedScales}
\Delta_\text{med}(t) = 1 - \exp\left[- \frac{1}{12} r_\perp^2(t) \Qs^2(t) \right] \,,
\eeq
where $r_\perp(t) = \theta_{12}t$ and $\Qs^2(t) = \hat q t$. The medium characteristics are all encoded in the parameter $\hat q$, discussed at length in Sec.~\ref{sec:MediumProp}. The function in Eq.~(\ref{eq:DeltaMedScales}) behaves as an order parameter describing two characteristic regimes and is therefore referred to as a `decoherence parameter'.
\begin{figure}[t]
\centering
\includegraphics[width=0.35\textwidth]{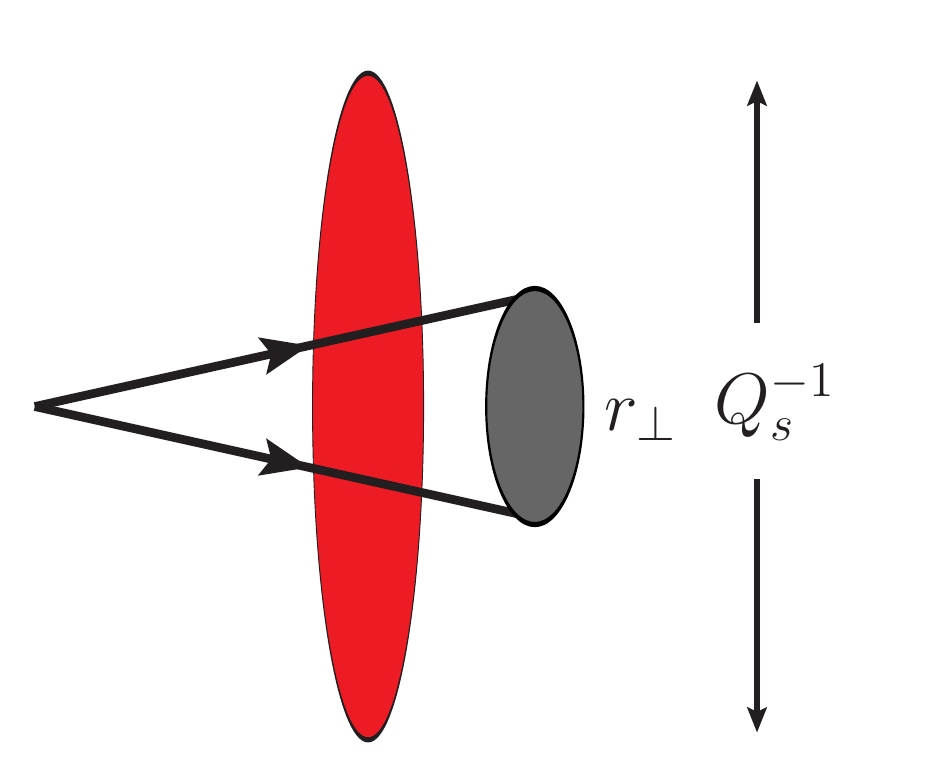}\hspace{1cm}
\includegraphics[width=0.28\textwidth]{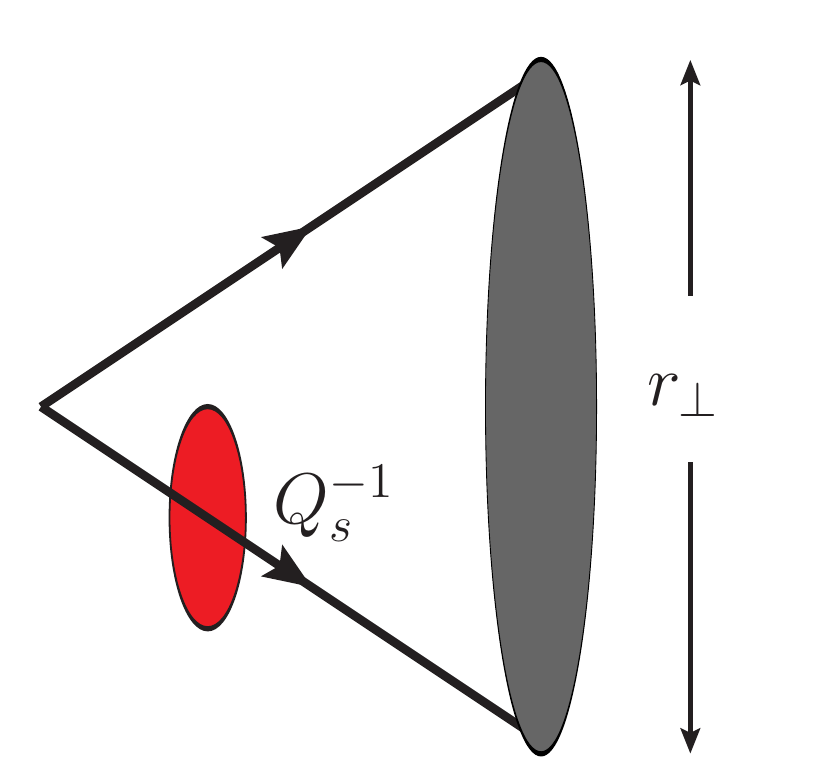}
\caption{The characteristic regimes of radiation in media: the `dipole` regime, $r_\perp \ll Q_s^{-1}$ (left) and the `decoherence` regime, $r_\perp \gg Q_s^{-1}$ (right). Figure taken from Ref.~\citen{MehtarTani:2011gf}.}
\label{fig:regimes}
\end{figure}
In the two extreme limits, it reads
\beq
\Delta_\text{med}(t,0) = \begin{cases}  0 \,, \quad r_\perp \ll \Qs^{-1} \;\text{(coherence)} \,,
\\ 1 \,, \quad r_\perp \gg \Qs^{-1} \;\text{(decoherence)}  \,,\end{cases}
\eeq
which possess a simple geometrical interpretation, see Fig.~\ref{fig:regimes}. In the former case, sometimes called the `dipole` regime, the pair remains correlated after propagating for a time $t$. In this case one can expand the exponent in Eq.~(\ref{eq:DeltaMedScales}) to obtain $\Delta_\text{med} \sim r_\perp^2$, which is the characteristic behavior of {\it color transparency}\cite{MehtarTani:2010ma,MehtarTani:2011tz,MehtarTani:2011gf}. In the latter case, called the `decoherence` regime, the pair is resolved by the medium interactions. In particular, this implies that all interferences between the particles are suppressed and the two particles behave as independent color charges.

\begin{figure}
\centering
 \includegraphics[width=0.5\textwidth]{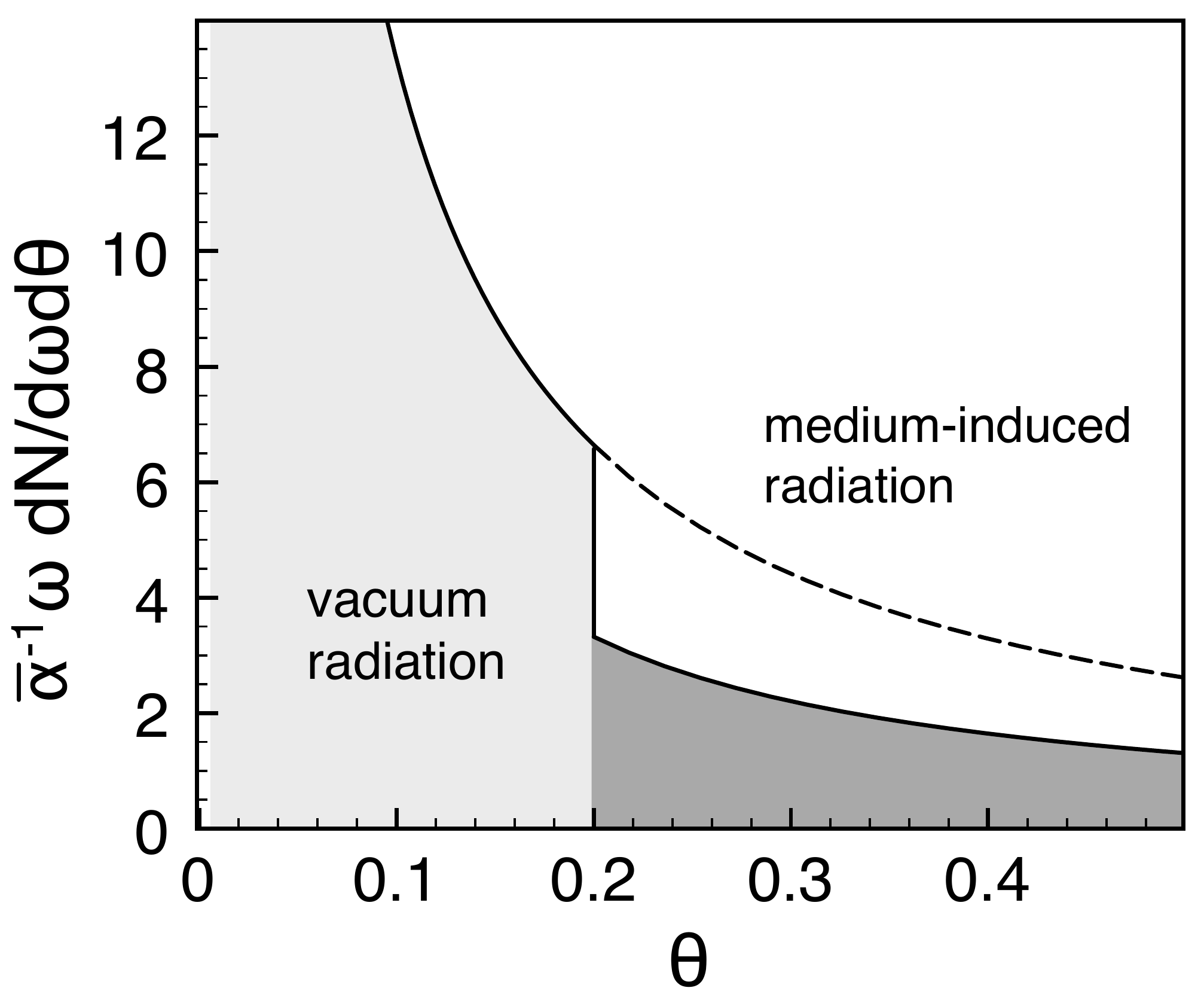}
\caption{The soft gluon emission spectrum off the quark constituent of a singlet antenna with opening angle $\theta_{12} = 0.2$, according to Eq.~(\ref{eq:SpectrumCoherentSoft}), in the presence of a medium with $\Delta_\text{med} = 0.5$ (solid line). Here $\bar \alpha \equiv \alpha_s C_F/\pi$. On average vacuum radiation is confined within $\theta < \theta_{12}$, while the medium-induced radiation is radiated at $\theta > \theta_{12}$. The limit of opaque medium, given by $\Delta_\text{med} = 1$, is marked by the dashed line. Figure taken from Ref.~\citen{MehtarTani:2011tz}.}
\label{fig:DecoherenceSoft}
\end{figure}

The role of color decoherence is most easily discerned in the soft sector where medium-induced gluon radiation is suppressed. Let us, for the purpose of illustration, presently focus on this regime. In this limit, all medium effects enter via a multiplicative factor $\Delta_\text{med}$, defined in Eq.~(\ref{eq:DeltaMedScales}), and the coherent spectrum off one of the antenna constituents, i.e. the spectrum in the presence of the charged carried by the other constituent\footnote{For a definition see Ref.~\citen{Dokshitzer:1991wu} and \citen{MehtarTani:2011gf,MehtarTani:2012cy}.}, after integrating over the azimuthal angle, can be written as
\beq
\label{eq:SpectrumCoherentSoft}
\left. dN \right|_\text{soft, coh} = \frac{\alpha_s C_R}{\pi} \frac{d\omega}{\omega}\frac{d\theta}{\theta} \Big[\Theta\left(\theta_{12} - \theta \right) - \Delta_\text{med}\Theta\left(\theta - \theta_{12} \right) \Big] \,,
\eeq
where $\omega$ and $\theta$ are the energy and angle of the emitted gluon, respectively. These features are illustrated in Fig.~\ref{fig:DecoherenceSoft}. In the absence of a medium, $\Delta_\text{med}\to0$, we recover the pure vacuum spectrum, which is explicitly constrained to angles smaller than the opening angle of the pair. This reflects the property of angular ordering.\cite{Dokshitzer:1991wu}
For a finite medium density, $\Delta_\text{med} > 0$, the medium-driven component occurs at large angles and is suppressed inside the antenna.\cite{MehtarTani:2010ma} In the limit of a completely opaque system, $\Delta_\text{med}\to 1$, the interferences are completely washed out and the soft emissions in the presence of a medium reduces to independent radiation off the quark and antiquark, as if they were radiating in the vacuum. From this simple picture we see that one of the chief implications of the onset of decoherence is that it marks the breakdown of the characteristic angular-ordering features of the jet as it propagates through the medium.

Correspondingly, the same condition of decoherence determines how the medium `resolves' the effective charge that subsequently will loose energy via induced radiation, described in detail in Sec.~\ref{sec:MediumInducedRad}.\cite{MehtarTani:2011gf,MehtarTani:2012cy} For higher gluon energies, following the above discussion, the interference diagrams also involve contributions from purely medium-induced gluons. The emerging picture can be summarized in the following simple principle,
\begin{center}
\begin{tabular}{ccl}
$\Delta_\text{med} = 0$ & $\rightarrow$ & medium-induced radiation off the total charge, \\
$\Delta_\text{med} = 1$ & $\rightarrow$ &medium-induced radiation off the constituent charges.
\end{tabular}
\end{center}
For instance, in case of an antenna in a color singlet configuration, e.g. originating from the splitting of a virtual photon $\gamma^\ast \to \qqb$, no radiation is induced by the medium in the `coherent` regime. Consequently, we observe that the decoherence of the color correlation between partons propagating through the medium also determines how they loose energy via radiative mechanisms.\footnote{Note that the same coherence effects are at play for purely elastic interactions in the medium.\cite{Neufeld:2011yh} }

The dynamics encoded in the decoherence parameter can also be translated in terms of a characteristic decoherence time,
\beq
\label{eq:DecoherenceTime}
\td \equiv \left(\frac{1}{ \hat q \,\theta^2_{12}} \right)^{1/3} = \left( \tform\, \tind^2 \right)^{1/3} \,,
\eeq
where in the last step we multiplied both numerator and denominator by the the energy carried by the antenna constituents. The timescales $\tform \sim 1/E \theta^2_{12}$ and $\tind$, see Eq.~(\ref{eq:HeuristicRadTime}), are the familiar branching times in vacuum and medium, respectively. Here, they both relate to the timescales over which the {\it antenna itself} is formed. This point brings us to two crucial insights. For highly energetic, formally $E\to\infty$, and strongly collimated probes, $\theta_{12} \ll 1$, the timescales for formation and decoherence are strongly separated, i.e., $\tform \ll \td$. Such configurations can only be realized by vacuum emissions. Secondly, for radiation which is characterized by $\tform = \tind$, we observe that $\tform = \td$. Since this type of radiation typically is distributed at large angles away from the jet direction, see Sec.~\ref{sec:MediumInducedRad}, we will not discuss it further in this Section.

Since the description of jet fragmentation is governed by the relevant scale, or ordering variable, it will be helpful to rephrase the physics of decoherence in the same language. Let us for the time being consider global scales, i.e., involving the total length of the medium.\footnote{We will denote $r_\perp \equiv r_\perp(L) = \theta_{12} L$ and $\Qs \equiv \Qs(L) = \sqrt{\hat q L}$.} Thus, besides the running scale of the intrinsic jet evolution, in the presence of a medium a novel hard scale given by
\beq
\label{eq:MediumHardScale}
Q_\text{med} = \max \left(r_\perp^{-1}, \Qs \right) \,,
\eeq
which determines the range (in transverse momentum) of the influence of the medium, will come into play. In particular, radiation with $k_\perp > Q_\text{med}$ is indeed coherent. This condition is met in two distinct cases.

The first case concerns {\it large-angle radiation}. If we consider the ordering features of radiation at angles $\theta > \theta_{12}$, the discussion above describes how angular ordering is spoiled for soft gluon radiation by the medium decoherence effects, see Eq.~(\ref{eq:SpectrumCoherentSoft}). For finite gluon energies, the maximal angle of radiation off the constituents of the pair becomes $Q_\text{med}/\omega$.\cite{MehtarTani:2011gf,MehtarTani:2012cy} This angle replaces essentially the opening angle of the pair for subsequent emissions. In the `decoherent` regime, where $Q_\text{med} = \Qs$, this angle agrees with the maximal angle of medium-induced radiation which provides a subtle consistency check: the system cannot transport its color charge to larger angles.

In a complementary fashion, coherence effects are also important for very {\it collinear radiation}. As pointed out above, these configurations are characterized by $\tform \ll \td$. Due to the ordering features of the vacuum shower, this opens up the possibility for multiple emissions within the window delimited by $\td$ which remain color coherent. This applies generally to hard and collinear radiation inside a jet, specifically all modes of the jet which meet the criterion for coherence.\cite{CasalderreySolana:2012ef} This possibility is particularly appealing in the context of high-$p_T$ jets. The presence of a hard scale due to the medium therefore converts to the following angular condition. All jet sub-structure contained within the critical angle which, in the approximations used throughout, reads
\beq
\theta_c = \left(\hat q L^3 \right)^{-1/2} \,,
\eeq
remains unresolved by the medium.\cite{CasalderreySolana:2012ef} Analogously to the antenna case, resolved jet substructures, i.e. separated at angles larger than $\theta_c$, evolve independent of each other and are not anymore forming a color coherent structure. Of course, this effective number of resolved subjets also depends on the specific fragmentation property of a given jet. It follows that the jet evolution on angles smaller than $\theta_c$ is completely vacuum-like and, consequently, one should expect that the distribution of these fragments scales like the vacuum distribution. Secondly, the unresolved system looses energy {\it coherently} with a rate as in Eq.~(\ref{eq:BDMPSrate}) proportional to the charge contained within $\theta_c$.

For realistic situations it was noted\cite{CasalderreySolana:2012ef} that a core containing most of the jet energy remains unresolved by the medium. This makes it a particularly appealing probe of the transport parameter $\hat q$ of the medium. Secondary substructures carry typically smaller energy fractions and are expected to be influenced quite violently by medium effects. In addition to the effects on  final-state (time-like) showers, color decoherence is also paramount when studying processes which involve several color charges, e.g., observables which become important at NLO, see Sec.~\ref{sec:JetQuenchingPhenomenology} for a further discussion. One particular example are multi-jet events, which in hadronic collisions typically involves the interplay of space-like and time-like emissions.\cite{Abe:1994nj,Abbott:1997bk} An interesting prospect would be to observe the modification of their coherent structure in the presence of a medium.\cite{Armesto:2012qa}

As discussed above, decoherence in course of subsequent branching leads to the disturbance of the color flow within the jet. This influences the later stage of the showering and affects the particle distributions created in the course of hadronization.\cite{Beraudo:2011bh} We emphasize that the established models of this non-perturbative process contain the basic elements of color coherence. That interactions of an outgoing probe with soft gluons from the background `reorganize` the color flow of the event has also been realized in the context of diffractive deep inelastic scattering.\cite{Edin:1995gi,Pasechnik:2010cm} Recently, it was noted that constituents of a jet that have interacted with the medium could end up in a configuration where they are color-connected with the medium excitations, rather than to other jet fragments.\cite{Beraudo:2011bh} These modifications have been calculated up to second order in medium density and also checked numerically with existing event generators.\cite{Beraudo:2012bq} This effect causes the formation of strings or clusters with typically larger invariant mass than for unperturbed color flow as in vacuum, leading to larger multiplicity of soft particles and consequently to overall softer spectra and a characteristic high-$p_T$ depletion.

\section{Phenomenology of jet quenching}
\label{sec:JetQuenchingPhenomenology}

\begin{quote}{\it
We have reviewed how decoherence of the jet governs its fragmentation pattern in the medium and determines the rate of coherent energy loss in some limiting cases. Based on these insights, we try to assess the present status of jet quenching phenomenology and the challenges ahead. The wealth of high-quality experimental data can help in pinning down the relevant effects.
}\end{quote}

\subsection{Modeling the shower}
\label{sec:JetQuenchingPhenomenology_Models}

The focus of this review is the physics of high-$p_T$ probes, and jets in particular, which traverse the complex state created in the wake of heavy-ion collisions. So far we have reviewed some of the chief elements related to parton branching, such as accumulation of transverse momentum and color decoherence, but the theoretical framework unifying these various aspects and allowing for a well-controlled comparison with data is yet to be established. A key issue currently under investigation is the space-time picture of the branching process and, in particular, how collinear and medium-induced radiation intertwines in the shower. For phenomenological applications one currently employs working models which are based on reasonable assumptions within a certain window of kinematical validity. Considering the limited space available, here we will only comment on a few key issues.

The single-gluon emission spectrum in a static medium defines a clear theoretical setup --- often referred to as the `QCD brick problem' for jet quenching.\cite{Armesto:2011ht} The main input to this calculation is the medium interaction potential containing the thermal effects and the mean free path, discussed in Sec.~\ref{sec:MediumProp}. In the high-energy limit this allows to compute, strictly speaking, the energy spectrum of medium-induced gluons, albeit numerically.\cite{CaronHuot:2010bp} Instead, most model implementations employ well-controlled approximations, such as the `harmonic oscillator` approximation discussed extensively above, which allow for analytical solutions. In this case, care has to be taken to correctly incorporate the limits of applicability and estimate errors. For instance, for a fixed value of the quenching factor, defined in Eq.~(\ref{eq:QuenchingFactor}), various implementations report a wide discrepancy of the resulting medium parameters, mainly $\hat q$.\cite{Horowitz:2009eb,Armesto:2011ht} This reflects the theoretical uncertainties inherent in the employed approximations.\footnote{Similar studies, including an expanding medium, have also been documented in Refs.~\citen{Bass:2008rv,Armesto:2009zi}.} These uncertainties should be systematically improved upon. In particular, finite-energy corrections can also play a significant role for quantitative estimates.\cite{Ovanesyan:2011xy,Ovanesyan:2011kn,Majumder:2012sh} Recently, a numerical Monte Carlo procedure has been formulated to correctly reproduce the LPM effect in QCD.\cite{Zapp:2011ya}

Due to the large phase space for radiation for high-energy probes, the second great challenge is to implement multiple emissions in the medium. It is useful to group the current working models into two broad categories set apart by the choice of evolution variable for the shower. 

On one hand, the constant medium-induced rate, Eq.~(\ref{eq:BDMPSrate}) and the discussion below, seems to indicate a strict {\it time ordering} of medium-induced radiation. The branching in medium is then described using the rate equation Eq.~(\ref{eq:RateEquationSoft}), generalized to account for energy-momentum conservation and including mixing of partonic species. Implementations of this scenario usually assume that the input partonic distributions can be taken as vacuum.\cite{Schenke:2009gb,Qin:2010mn}

On the other hand, assuming that the initial virtuality of the jet is much larger than any of the scales from the medium, one can treat the medium-induced splitting processes as a small perturbation on top of the vacuum evolution {\it ordered in virtuality} which resums collinear emissions, see Sec.~\ref{sec:VacJets}. This approach is also appealing from the point of view of Monte-Carlo implementation.\cite{Zapp:2008af,Renk:2008pp,Zapp:2012ak,Armesto:2009fj} One can argue heuristically that this holds for splittings where the drop in virtuality exceeds the amount of transverse momentum accumulated from the medium during the formation time,\cite{Majumder:2010qh} $k_\perp^2 \gtrsim \sqrt{\hat q \omega}$, where $k_\perp$ is the transverse momentum of the gluon created in the splitting. This can be accommodated in various ways. Here we bring forward one particular prescription where a medium-modified splitting kernel is defined\cite{Guo:2000nz,Wang:2001ifa,Polosa:2006hb,Majumder:2009ge}
\beq
\label{eq:MMSplitingKernel}
P_{ij}^\text{tot}(z) = P_{ij}(z) + \Delta P_{ij}^\text{med} (z; \,\hat q, L, \ldots) \,,
\eeq
which generalize the standard Altarelli-Parisi\cite{Altarelli:1977zs} ones. The medium-modified part, second term in Eq.~(\ref{eq:MMSplitingKernel}), is again related to the rate of medium-induced gluons, see e.g. Eq~(\ref{eq:BDMPSspectrum}) and Refs.~\citen{Guo:2000nz,Wang:2001ifa,Majumder:2009ge} for other deriviations. Note that it typically involves the {\it global} scales involving the full medium length $L$, setting it apart from the rate equations discussed earlier on, which are {\it local} in time. The full splitting function is used as input to the standard DGLAP evolution equations of the hadron fragmentation function, see Sec.~\ref{sec:VacJets}. In the limit of soft gluons, $z \ll 1$, this approach also leads to nearly independent emissions.\cite{Armesto:2008qh}

The elements discussed so far can be improved systematically, at least in some limiting cases, within the framework of perturbation theory. Additionally, the influence of several {\it non-perturbative} effects are necessary for a realistic comparison with data and the complexity of the problem at hand grows rapidly. This purely phenomenological input, from the point of view of jet quenching, includes the modeling of the initial nuclear geometry and space-time picture of the plasma evolution, typically taken from hydrodynamical models. Considering jet quenching as a tomographic probe of the QGP, these effects can play an important role when comparing to experimental observables.\cite{Renk:2011aa} The fate of the produced color charges in course of medium-induced radiation\cite{CasalderreySolana:2010eh} and their back-reaction on the plasma evolution is still under discussion. Finally, as discussed in Sec.~\ref{sec:CoherenceMedium}, medium effects on hadronization can also influence the measured hadronic spectra.\cite{Beraudo:2012bq} An additional complication arises for jet observables which demand subtle background subtraction procedures for an apples-to-apples comparison to experimental data.\cite{Cacciari:2010te,Apolinario:2012cg,Cacciari:2012mu,deBarros:2012ws} 

Presently, state-of-the-art calculations of jet quenching effects usually involve numerical event generation.\cite{Wang:1991xy,Lokhtin:2008xi,Zapp:2008af,Renk:2008pp,Armesto:2009fj,Schenke:2009gb,Buzzatti:2011vt,Zapp:2012ak} Although many codes are used by several groups for comparison with experimental data, only a few are properly documented in detail.

\subsection{Lessons from the data}
\label{sec:JetQuenchingPhenomenology_Data}

With the advent of the heavy-ion program at the LHC the physics of heavy-ion collisions has entered a new, high-energy regime.\footnote{Since the RHIC heavy-ion results have already been reviewed elsewhere, see e.g. Ref.~\citen{Majumder:2010qh}, we will presently focus on the new results from the ALICE, ATLAS and CMS Collaborations at the LHC.} 
The unprecedented kinematical reach together with a large lever in collision energy compared to RHIC provides a wide range to test the physics of energy loss under widely different conditions.
Ample production of high-$p_T$ jets and the excellent detector capabilities allow for the first time precise experimental data on their modifications in various aspects. Without going into the details of jet reconstruction in heavy-ion collisions,\cite{Cacciari:2010te,deBarros:2012ws} the jet sample discussed below usually comprise high-$p_T$ jets reconstructed using the anti-$k_t$ algorithm\cite{Cacciari:2008gp,Cacciari:2011ma} with a relatively narrow cone size $R\sim$ 0.2--0.5 to minimize the effects of background fluctuations. For an overeview of the most recent developments together with references to unpublished results, see also Ref.~\citen{CasalderreySolana:2012bp}.

Just as for the inclusive hadron production, a key measurement is the inclusive spectrum of jets in heavy-ion collision. Due to the highly nontrivial procedure to extract the jets from a violently fluctuating background,\cite{Abelev:2012ej} these unfolded spectra are not yet published but many of the inherent uncertainties cancel out in the construction of the $R_{AA}$ modification factor Eq.~(\ref{eq:NuclearModificationFactor}) (or the equivalent $R_{CP}$ where central collisions are compared to peripheral ones). The measurements indicates robustly a suppression by a factor of two for jets with $p_T$ = 50--200 GeV/c\cite{:2012is} which is roughly consistent with the measurement of the suppression of high-$p_T$ charged hadrons.\cite{Aamodt:2010jd,Abelev:2012hxa} The suppression at the highest available $p_T$ is remarkably independent of the cone size except at $p_T$ = 50--100 GeV/c where a decreasing trend of the suppression is seen as one opens the jet cone.\cite{:2012is} These data, together with the path length dependence as, e.g., measured by the second azimuthal Fourier coefficient $v_2$ of high-$p_T$ jets, see Ref.~\citen{CasalderreySolana:2012bp}, could help in constraining the nature of energy loss for these energetic probes.

Another important observable is the di-jet energy imbalance.\cite{Aad:2010bu,Chatrchyan:2011sx,Chatrchyan:2012nia} Due to the surface bias of triggered probes one typically expects the sub-leading jet to traverse a significantly longer path in the medium than the leading one. Remarkably, while the di-jet data manifest a significant enhancement of energy imbalance, the system remains almost ideally back-to-back. In fact, in central collisions the sheer number of candidate di-jets that pass the required experimental cuts is strongly reduced. This would imply that energy is flowing out of the jet cones mostly as soft quanta, $p_T \leq$ 2 GeV/c, and this expectation is indeed confirmed by looking at very large angles away from to the di-jet axis.\cite{Chatrchyan:2011sx,Chatrchyan:2012nia} This dynamics put strong constraints on the mechanism of energy loss --- in particular, the mechanism behind the transfer of a significant energy into soft modes at large angles. These general features have also been confirmed using photon-jet correlations.\cite{Chatrchyan:2012gt}

The recent data also allows to study the jet substructure both in terms of longitudinal momentum fraction $z$ of jet constituents and in terms of their the angle $r$ away from the jet axis per observed jet,\cite{Chatrchyan:2012gw} thus providing a `tomography' of jet modifications. Due to the QCD dynamics one naturally expects the hardest fragments to be distributed narrowly around the jet axis while the soft components typically tend to extend up to large angles. Thus, these combined measurements confirm that the `core' of the jet, i.e. hard components close to the axis, are not strongly affected by the medium. This is generically anticipated due to the physics of coherence, see Sec.~\ref{sec:CoherenceMedium}. The soft components at the fringes of the jet, on the other hand, are enhanced compared to the vacuum expectation.\cite{Chatrchyan:2012gw}

The features of the experimental data follow generic expectations of radiative processes in the medium --- on one hand, the physics of coherence related to highly collimated and energetic substructures in jets and, on the other, the large-angle medium-induced spectrum which carry the imprint of the local medium scale.

\section*{Acknowledgments}
KT would like to thank N.~Su for helpful comments. YMT is supported by the European Research Council under the Advanced Investigator Grant ERC-AD-267258. JGM work was partly supported by the Funda\c c\~ao para a Ci\^encia e a Tecnologia (Portugal) under project CERN/FP/123596/2011. KT is supported by a Juan de la Cierva fellowship and by the research grants FPA2010-20807, 2009SGR502 and by the Consolider CPAN project. 


\end{document}